\documentclass[useAMS,usenatbib]{mn2e}
\usepackage{amsmath}
\usepackage{color}
\usepackage{ctable}
\usepackage{epsfig}
\usepackage{epstopdf}
\usepackage{float}
\usepackage{graphicx}
\usepackage{times}

\newcommand{\atlas}{ATLAS$^{\rm 3D}$}
\newcommand{\kms}{$\rm km\,s^{-1}$}

\newcommand{\Reff}{$R_{\rm eff}$}
\newcommand{\lambdaR}{$\lambda_{\rm R}$}
\newcommand{\Hbeta}{H$\beta$}
\newcommand{\alphafe}{$[\alpha/\mathrm{Fe}]$}
\newcommand{\Mgb}{Mg${\it b}$}

\def\aap{A\&A}  

\def\apjl{ApJ}  
\def\apj{ApJ}  
\def\apjs{ApJS}   
\def\nat{Nature}  
\def\mnras{MNRAS} 
\def\pasp{PASP}  
\def\aj{AJ}  

\newcommand{\placetab}{
\begin{table}
\caption{Best-fit stellar population models for the different components obtained via a $\chi^{2}$ fit with CvD SSP models with a x=1.8 IMF slope.}
\label{tab:SSP}
\begin{center}
\begin{tabular}{cccc} 
\hline
{\bf Component} & {\bf Age (Gyr)} & { \bf [M/H] } &  { \bf  [$\alpha$/Fe] } \\
\hline
 Central & 	$12.9^{+1.5}_{-1.8}$ &	 $+0.23^{+0.08}_{-0.05}$ &$+0.09^{+0.03}_{-0.01}$	   \\	
 Bulge &  $13.5^{+1.4}_{-1.4}$ &	 $-0.17^{+0.12}_{-0.1}$ &	 $+0.03^{+0.03}_{-0.01}$\\
 Disk & 	 $10.5^{+1.6}_{-2.0}$ &	 $-0.03^{+0.02}_{-0.1}$     &	 $+0.04^{+0.02}_{-0.01}$   \\      
\hline
\end{tabular}
\end{center}
\end{table}
}

\newcommand{\placefigfov}{
\begin{figure*}
\begin{center}
\includegraphics[height=13cm]{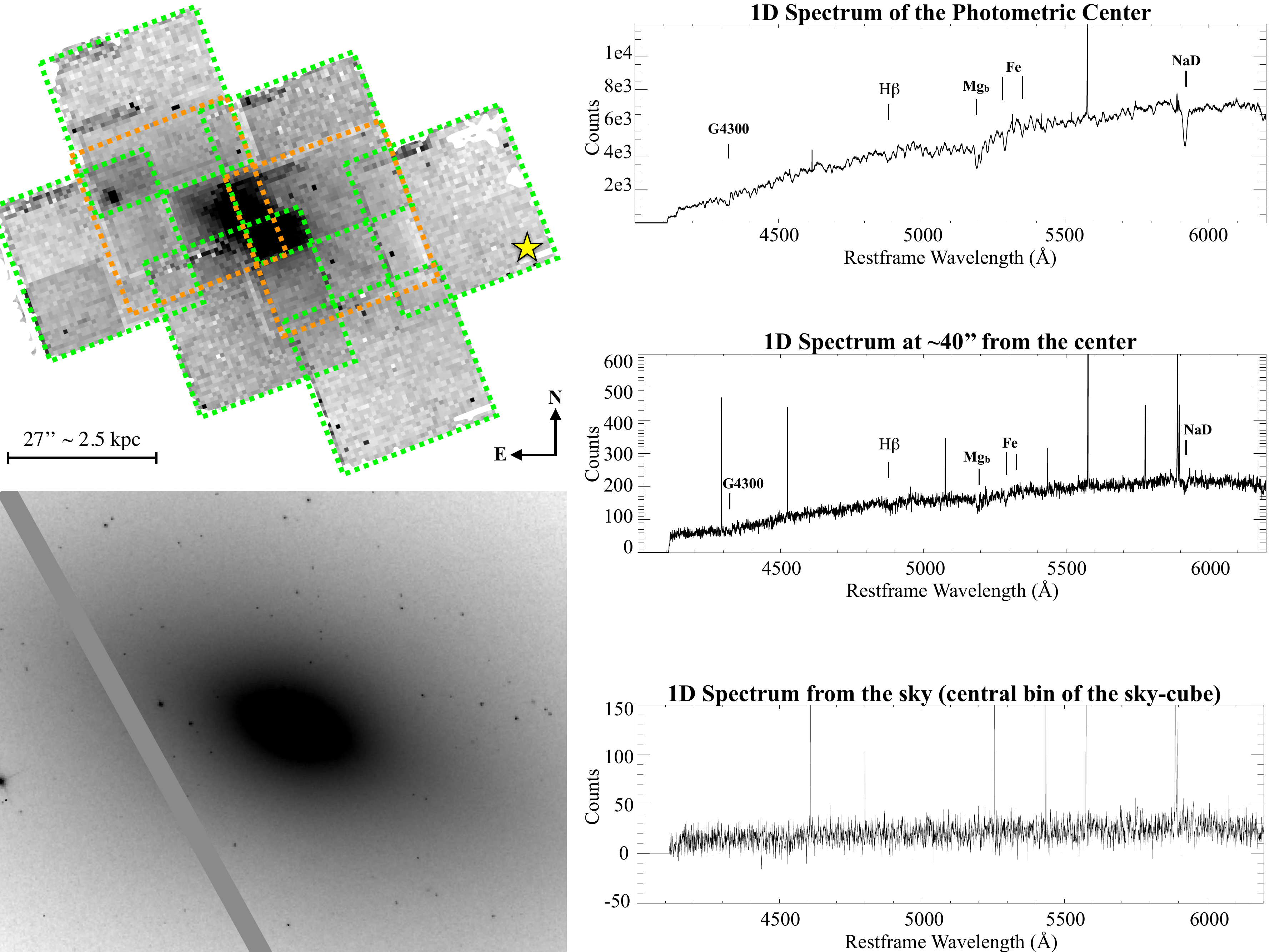}
\caption{{\sl Left-upper panel: } NGC~4697 final reconstructed field-of-view generated with the mosaic of all the on-target pointings, summing 
the fluxes in the datacube over the whole wavelength range. 
Each dotted square represents a single pointing of $27''\times27''$. Green cubes are data from the first run (P79), 
whereas orange cubes  are the two central pointings from the second one (P85). 
{\sl Left-bottom panel:} {\it HST} ACS image of the central region of NGC~4697 showed at the same spatial scale as the reconstructed VIMOS image. 
Blue isophotes (derived from the {\it HST} image) are shown on both images to check the validity and precision of the astrometrical calibration (see text for more details). 
{\sl Right panels:} 1D galaxy spectra extracted in the central spatial bin (top panel) and in a spatial bin 
at $\sim 40"$ from the galaxy center (middle panel), and 1D sky spectrum (bottom panel) extracted from the central bin of the skycube. 
The spaxel from which the middle-panel 1D spectrum has been extracted is highlighted on 
the reconstructed FOV by a yellow star. The spectra are plotted versus restframe wavelength and several 
spectral features of the galaxy such as  G4300, \Hbeta, \Mgb\, and Fe lines are marked. 
Sky emission lines have been not removed from the spectra but have been properly masked out during the fit procedures. }
\label{fig:fov}
\end{center}
\end{figure*}
}

\newcommand{\placefigmappe}{
\begin{figure}
\includegraphics[height=13cm]{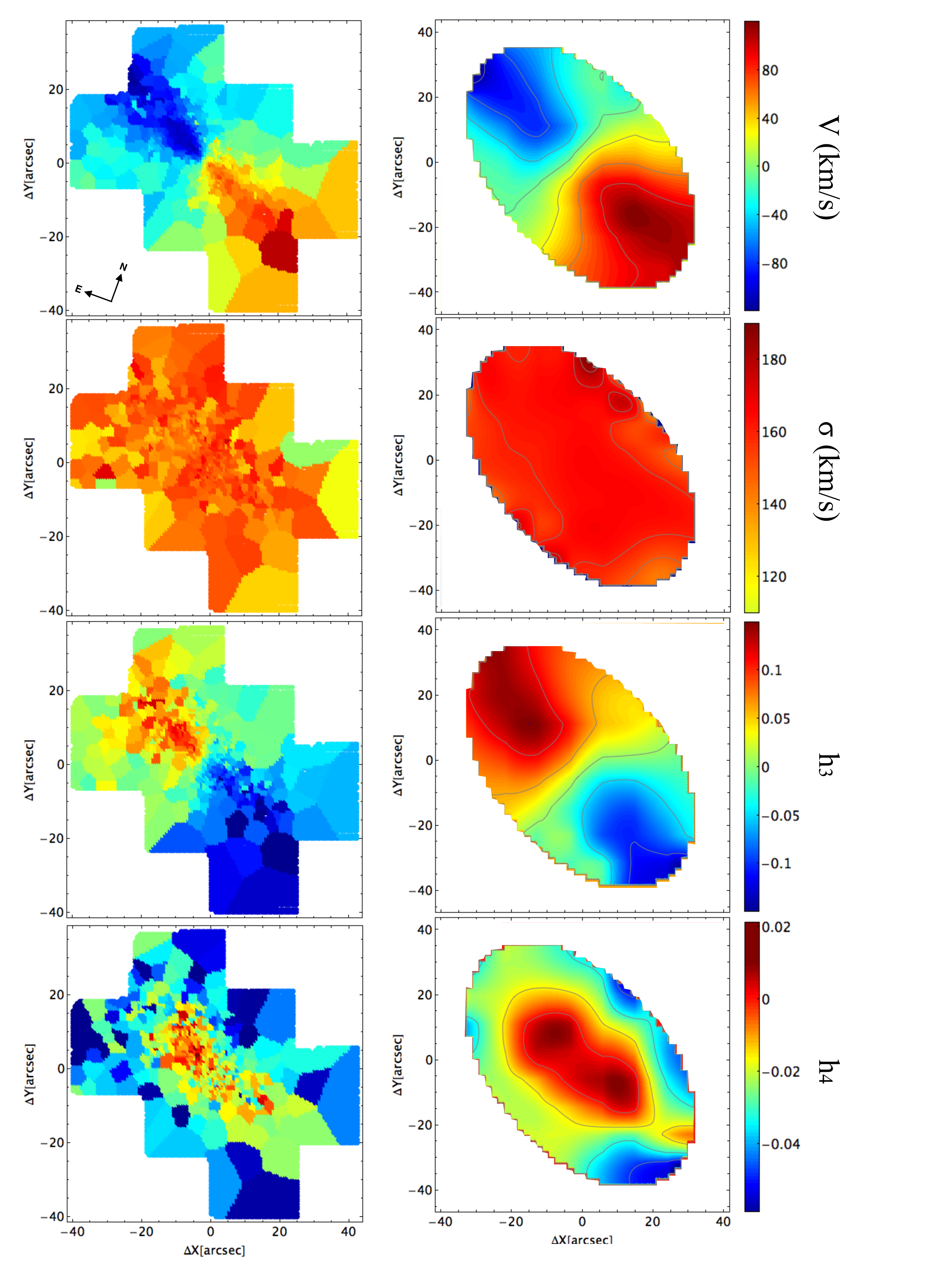}
\caption{NGC~4697  2D line-of-sight velocity distribution maps in a $80" \times 80"$ box. 
Panels on the left column show the un-smoothed  Voronoi-rebinned version, as derived in Sec.~\ref{subsect:voronoi}. 
Here the small scale features are visible. 
Panels on the right column show the  2D  maps smoothed by averaging the Voronoi values over a grid 
of 15$\times$15 macro bins and then masked with an ellipse mimicking the orientation and the 
axis ratio of the outermost isophote enclosing the VIMOS mosaic. Here the large scale pattern of the kinematical 
quantities is more appreciable. }
\label{fig:mappe}
\end{figure}
}

\newcommand{\placefigVsigma}{
\begin{figure}
\includegraphics[height=7.2cm]{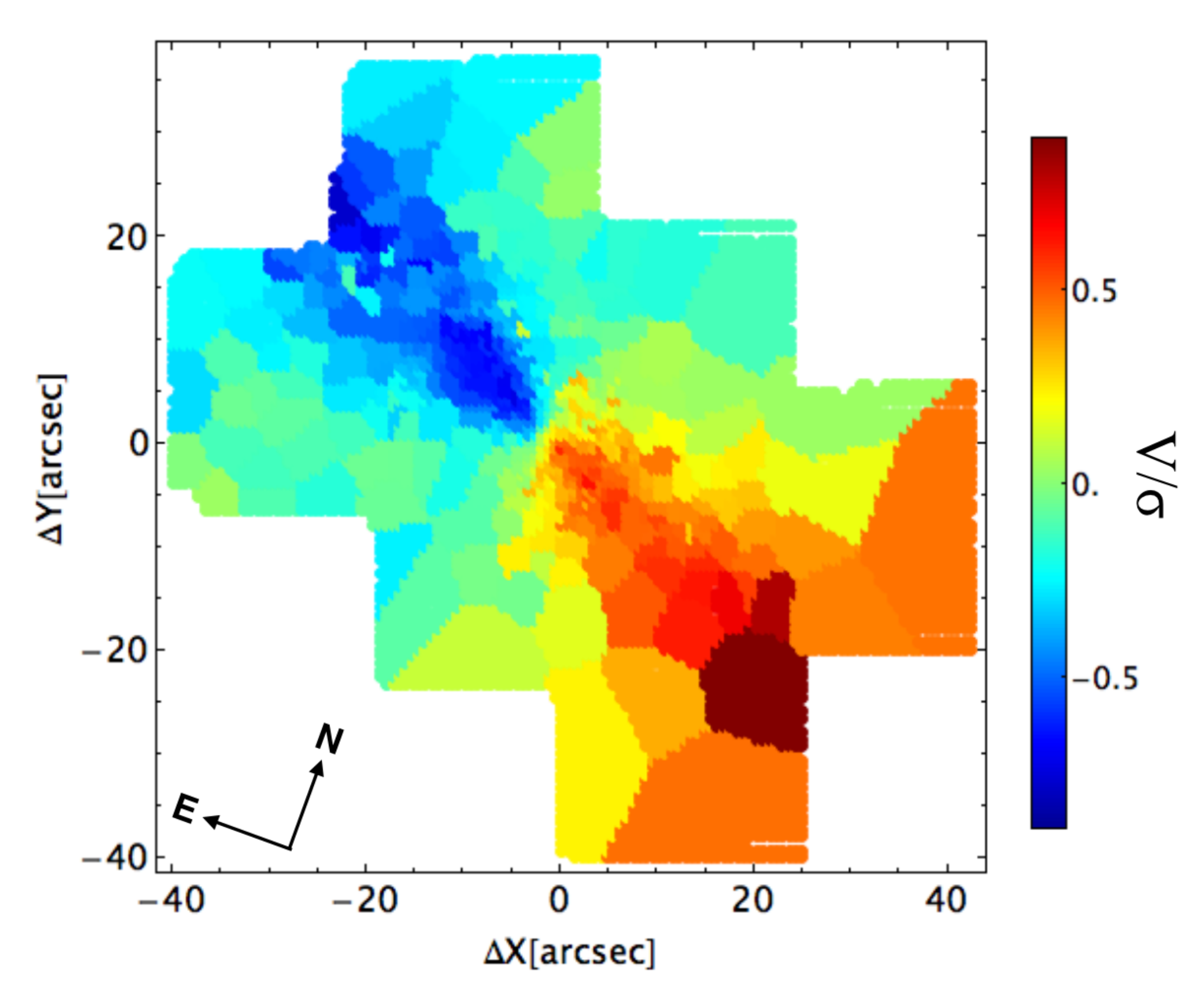}
\caption{{\sl Top panel}: $V/\sigma$  2D Voronoi map, quantifying the ratio of the ordered 
and random motion of NGC~4697.}
\label{fig:vsig}
\end{figure}
}

\newcommand{\placefigoneDprof}{
\begin{figure}
\includegraphics[height=11cm]{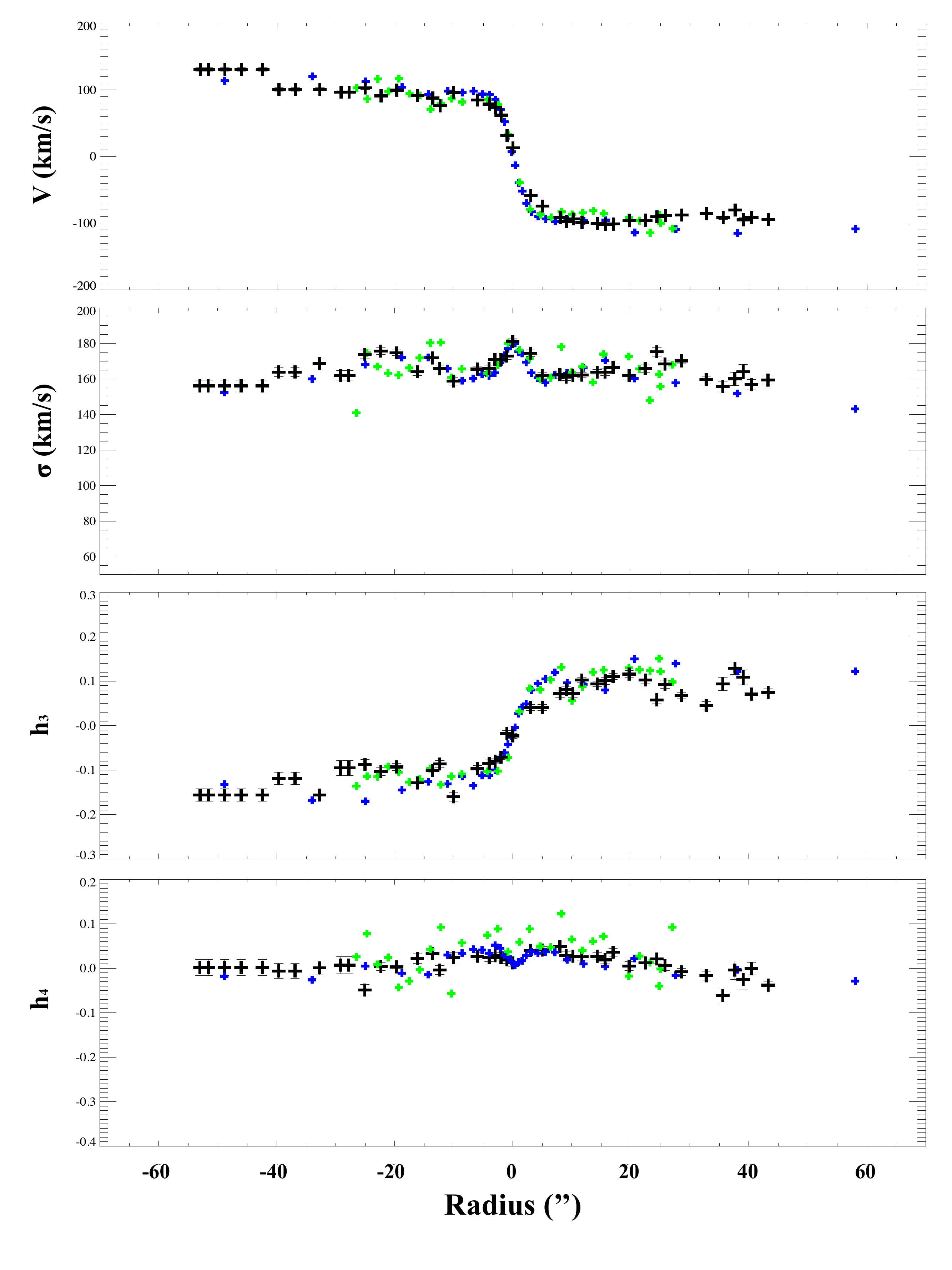}
\caption{NGC~4697 1D kinematic profiles along the major axis. Black points are data from this work, 
blue points are long-slit data from \citet{deLorenzi2008} and green points are IFU data from the \atlas survey, 
extracted along the major axis with the same procedure used to extract our points. A good agreement is found between the different studies. 
Uncertainties on the rotation and velocity dispersion are of the order or 10 \kms, while on the $h_{3}$ and $h_{4}$ 
parameter are of the order of 10 percent.}
\label{fig:delor}
\end{figure}
}

\newcommand{\placefigelemmaps}{
\begin{figure}
\begin{center}
\includegraphics[height=7.5cm]{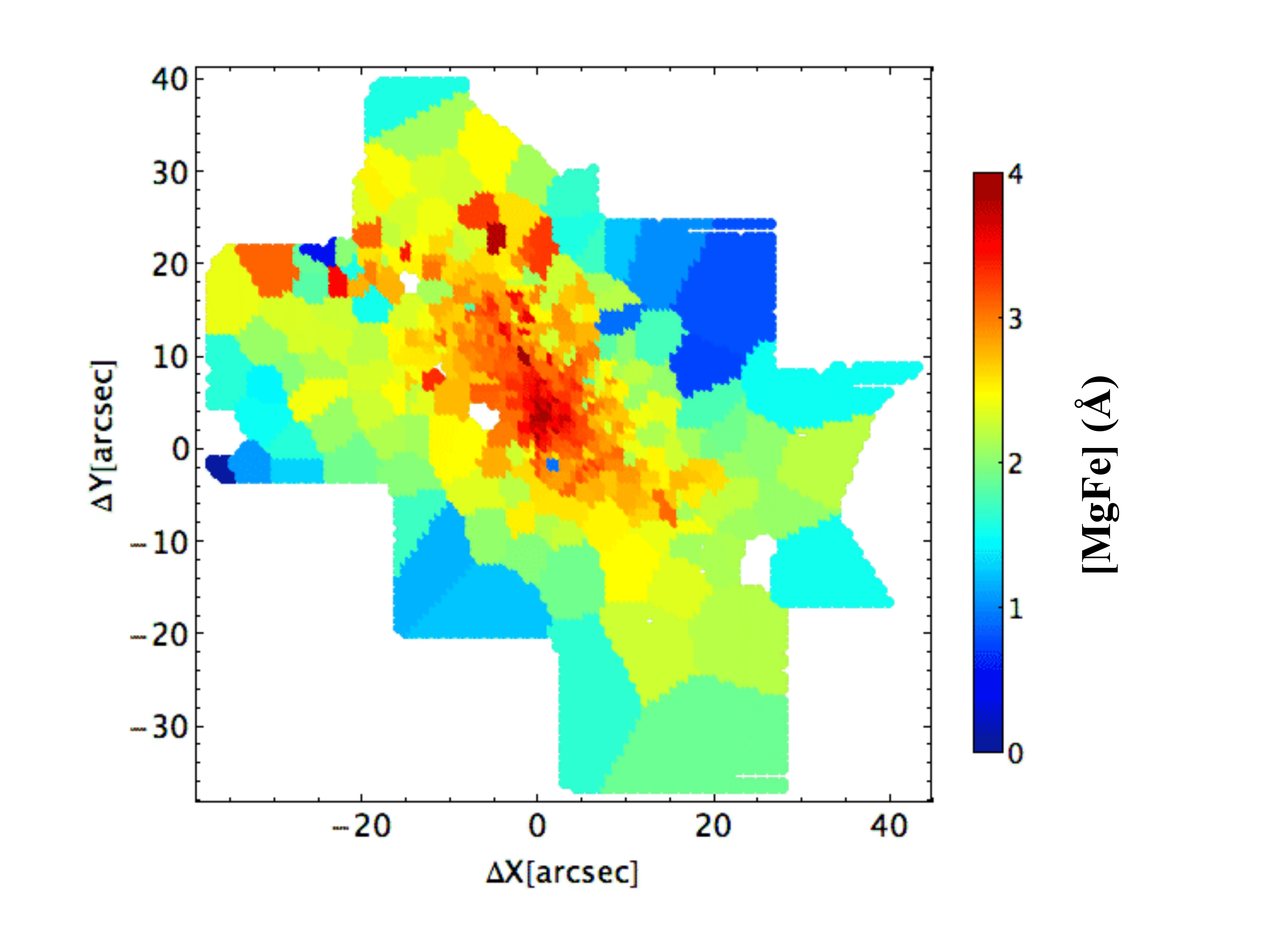}
\caption{NGC~4697 2D maps of the [MgFe] stellar absorption feature measured from the spectra
at a final resolution of $\sigma=350$ \kms. 
The map clearly shows that the stellar population of the disk is different 
from the stellar population of the bulge component of the galaxy. Areas within 
the disk are infact  more metal-rich (higher [MgFe] EWs).} 
\label{fig:element_maps}
\end{center}
\end{figure}
}

\newcommand{\placefigoneDspec}{
\begin{figure*}
\centering
\includegraphics[height=9cm]{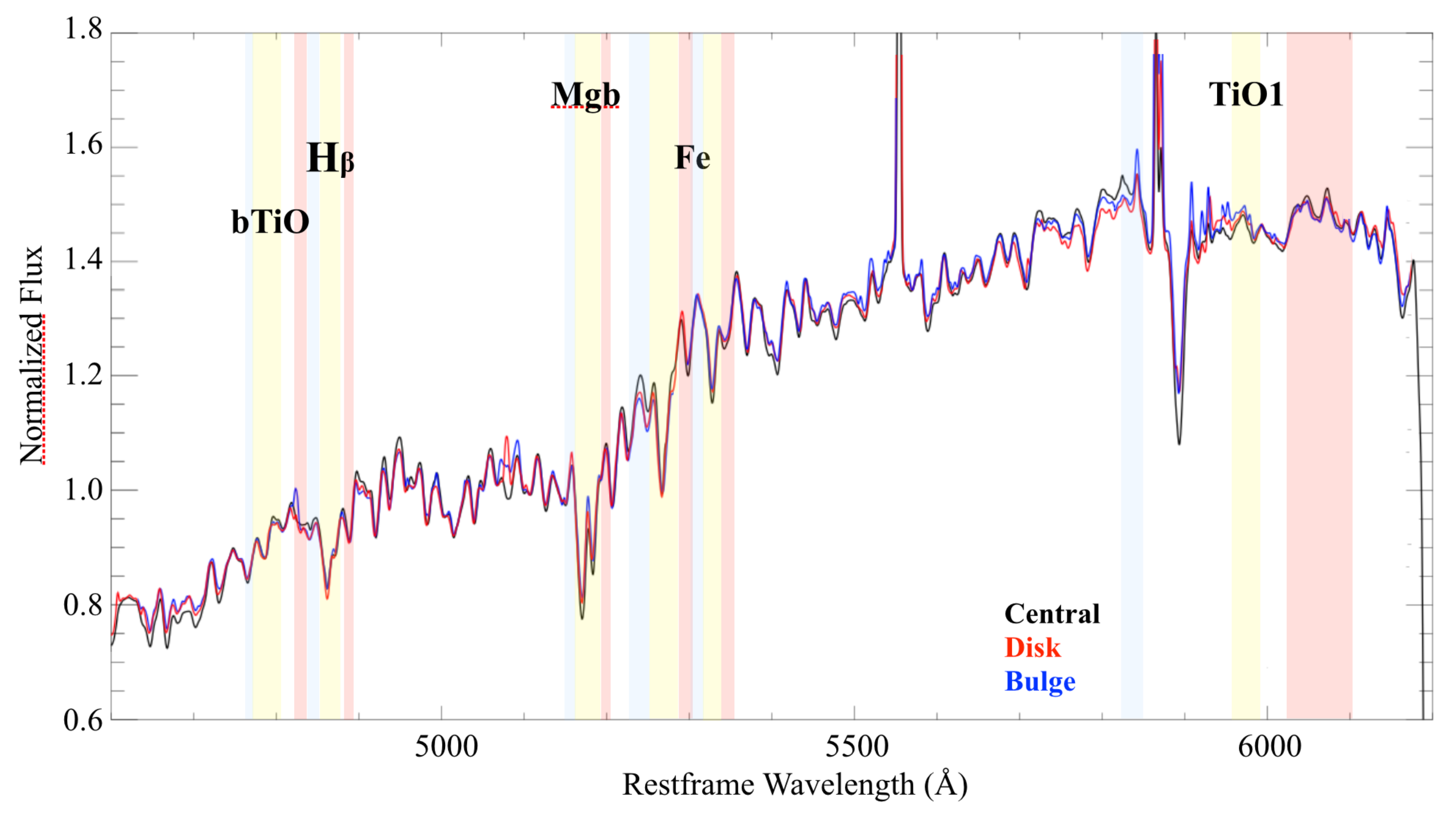}
\caption{1D spectra extracted from the bulge (blue), disk (red) and the central  $\sim$\Reff$/8$ region ($10$ arcsec, black). 
Yellow boxes show the band-passes of the line indices used to constrain the SP paramenters, 
while blue and red bands represent the blue and red continuum side-bands respectively.}
\label{fig:1dspectra}
\end{figure*}
}

\newcommand{\placefigstelpop}{
\begin{figure*}
\includegraphics[height=16 cm]{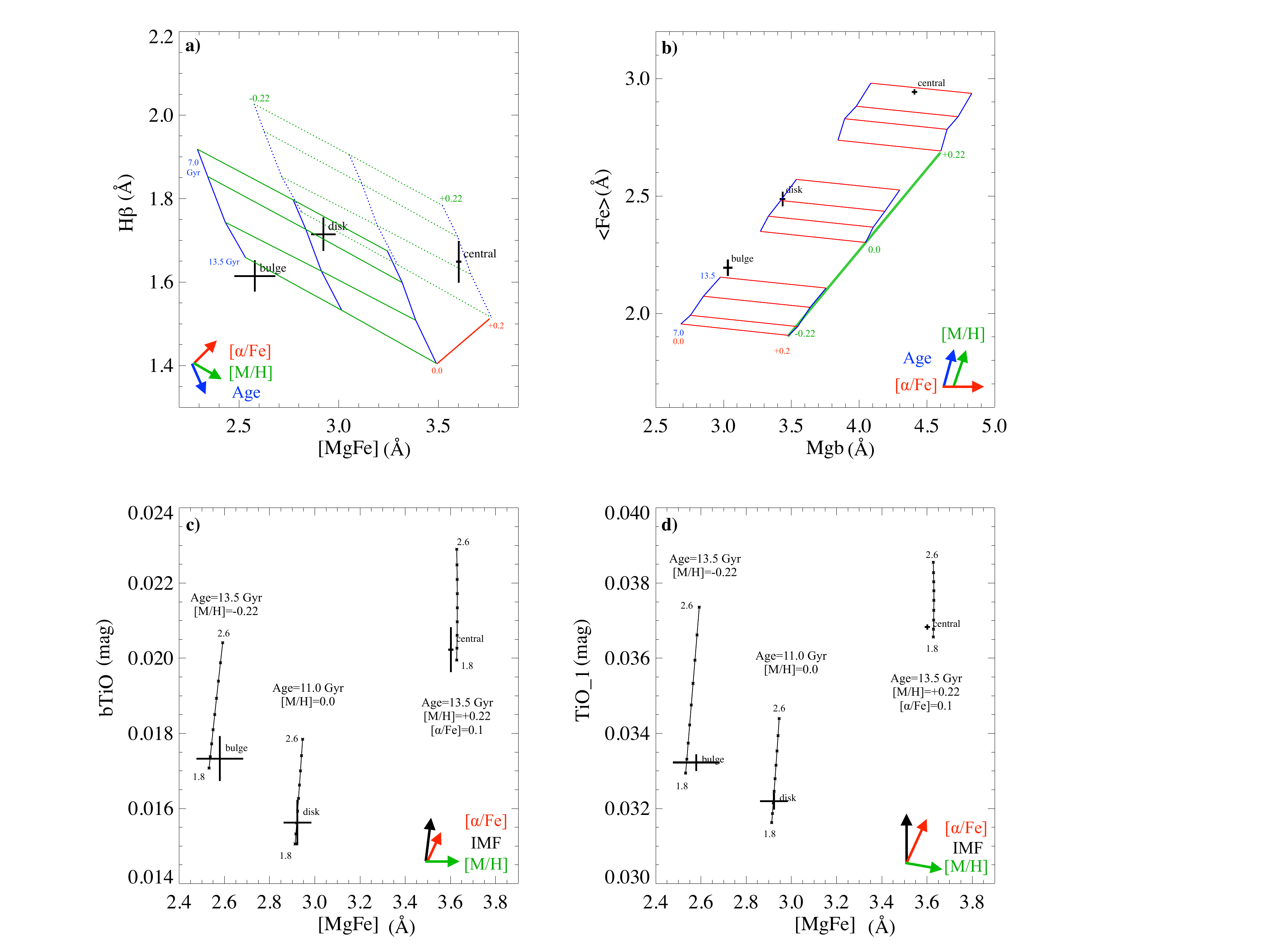}
\caption{Index-index plots of some of the classical stellar absorption features and 
of the bluer IMF-sensitive optical indicators defined in S+14 in the optical regime.  
Grids are CvD12 extended SSP models obtained in S+15 via the response function approach (see text for further details) . 
Blue lines show models with varying ages with values [7,9,11,13.5] Gyr,  
red lines are SSP models with varying \alphafe: [0.0,0.2,0.4] dex, green lines are 
SSP models with varying total metallicity  ($[M/\mathrm{H}] = [-0.22, 0.0, +0.22]$) 
and, finally, black lines are models with varying IMF slope from $x=1.8$ (Milky Way-like) 
to $x=2.6$ (being Salpeter $x=2.35$). 
Colored arrows in the corner of each panels show the direction of the variation of these three 
single parameters.  
Points with error bars are high S/N galaxy spectra extracted from the different regions 
by summing the Voronoi-rebinned spectra. Galaxies and models are convolved to a final 
common resolution of $350$ \kms\, before measuring index strengths. 
In the plots on the upper row (panels a and b),  IMF dependence is minimal 
(all models have x=1.8 here) and the age and abundances of the galaxy can be inferred. 
The IMF normalization of the galaxy can be instead inferred from panels c) and d), 
using SSP models with the metallicity and age constraints previously determined.  
Here, lines are CvD12 models with varying IMF slope in the range 1.8--2.6. 
}
\label{fig:index-index}
\end{figure*}
}

\newcommand{\placefiglast}{
\begin{figure*}
\centering
\includegraphics[height=13cm]{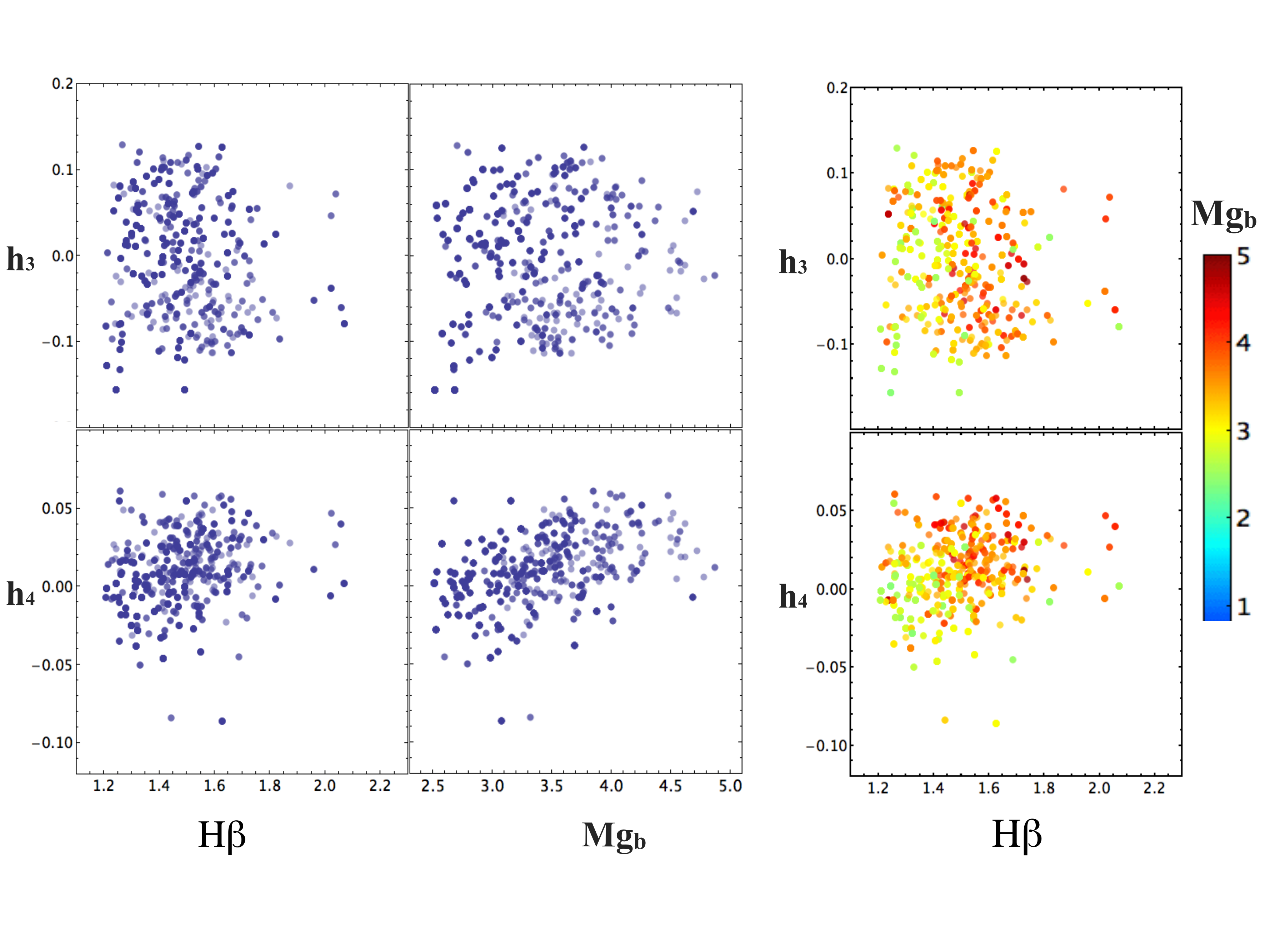}
\caption{
{\sl Left panels:} \Hbeta\, EWs versus the $h_{3}$ (top) and the $h_{4}$ (bottom) parameter. 
{\sl Middle panels:} \Mgb\ EWs  versus the $h_{3}$ (top) and the $h_{4}$ (bottom) parameter. 
In all panels, points represents values obtained from the macro bins of the smoothed maps 
and show a correlation between LOSVD properties and stellar population features. 
Larger \Hbeta\,  values roughly indicate younger ages, while larger \Mgb\, values indicate higher total metallicity 
(although the well-known age-metallicity degeneracy makes things a bit more complicated than that; e.g. \citet{Worthey1994}. 
{\sl Right panels:}  \Hbeta\, EWs versus the $h_{3}$ (top) and the  $h_{4}$ (bottom) parameter, 
color-coded by \Mgb\, EW values.  
Points with larger, positive $h_{4}$, aligned with the major axis (see Fig.~\ref{fig:simulh3h4}) 
have on average higher \Hbeta\, and  \Mgb . This is true also for points with larger  $h_{3}$ values (in absolute value) 
although the plot looks a bit noisier.  
This Figure highlights the agreement between kinematics and stellar population analysis both confirming that stars aligned 
with the major axis (where the disk dominates) are  younger and more metal-rich. }
\label{fig:h4hbeta}
\end{figure*}
}

\newcommand{\placefigsimul}{
\begin{figure*}
\includegraphics[height=12cm]{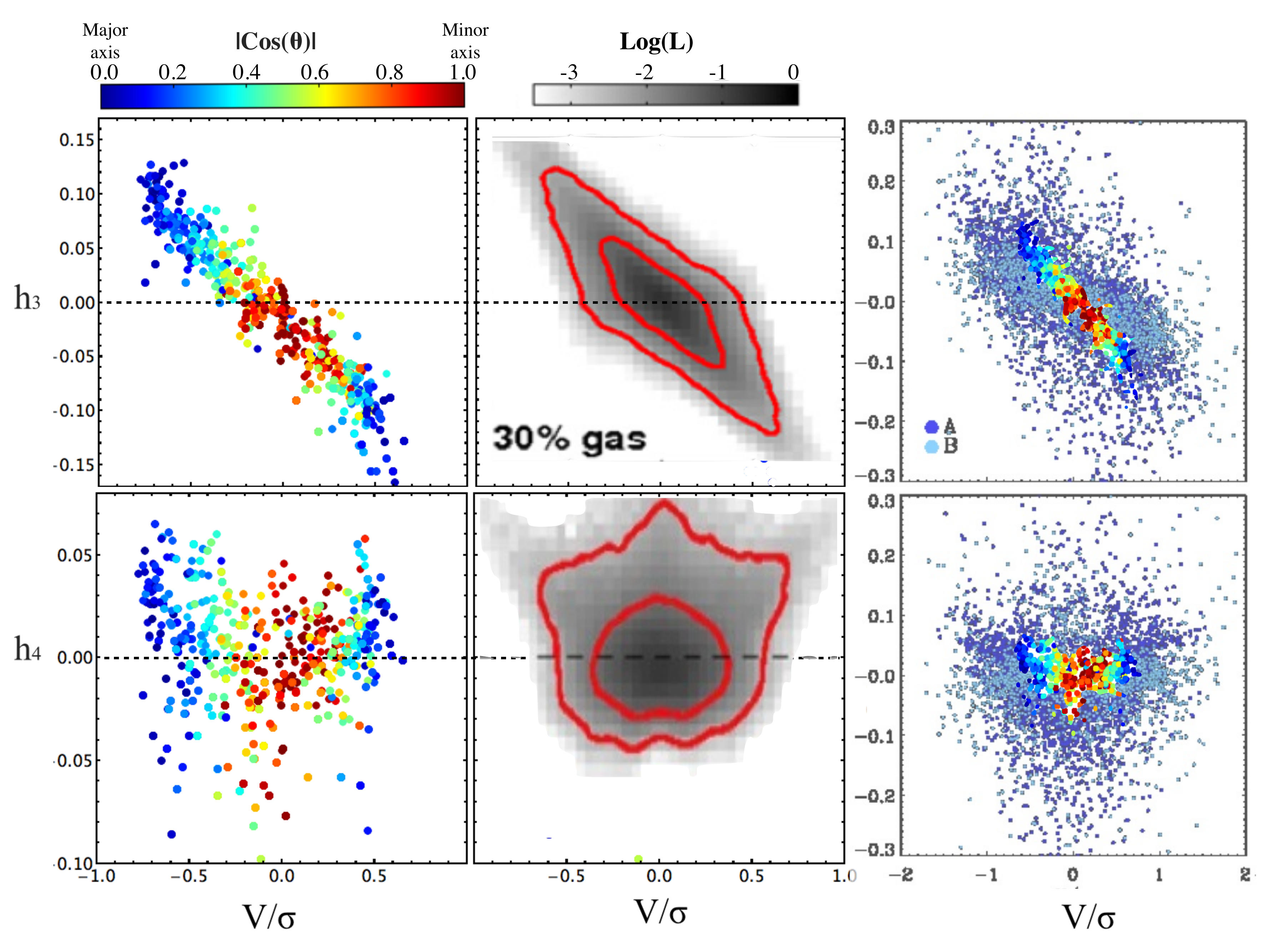}
\caption{{\sl Left panels:}The $h_{3}$--$V/\sigma$ (top) and the $h_{4}$--$V/\sigma$ (bottom) 
distributions from the VIMOS kinematic maps.  Points are color-coded according to the azimuthal 
deviation form the galaxy minor axis. 
{\sl Middle panels:} Comparison with predictions from 1:1 disk--disk merger simulations from H+09 
with a 30 percent gas fractions. The shading is proportional to the logarithm of the 
luminosity, whose scale is plotted in the bar on the top and the red contours contain 68\% and 95\% of the luminosity.  
{\sl Right panels:} Comparison with predictions of N+14 simulations for galaxies of class A and B. }
\label{fig:simulh3h4}
\end{figure*}
}

\newcommand{\placefiglambdar}{
\begin{figure}
\includegraphics[height=6.5cm]{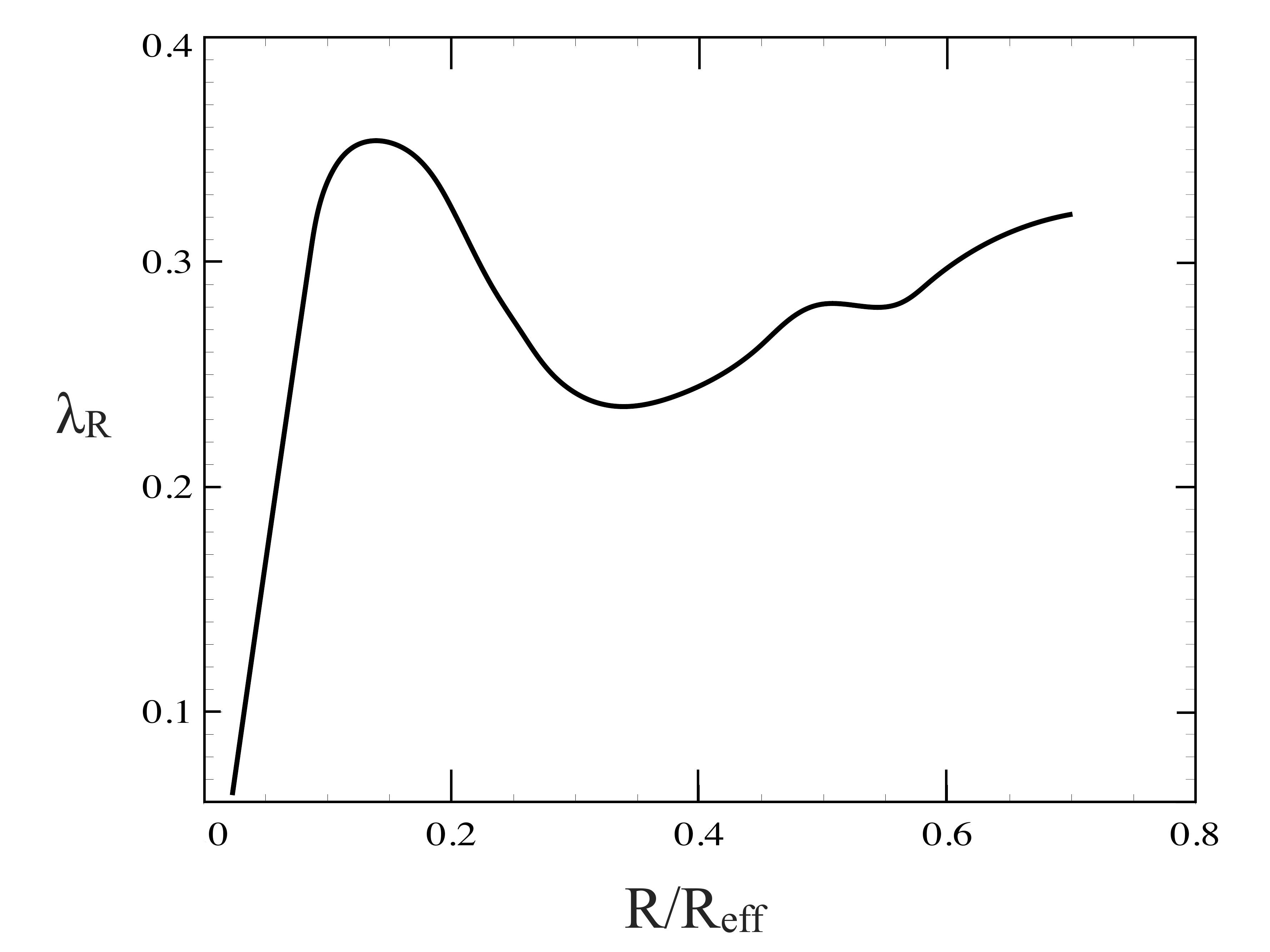}
\caption{ The \lambdaR\, parameter calculated using the formula of \citet{Emsellem2007} 
as a function of radius.}
\label{fig:lambdaR}
\end{figure}
}

\title[VIMOS mosaic of the ETG NGC~4697]{VIMOS mosaic integral-field spectroscopy of the bulge and disk of the early-type galaxy NGC~4697}
\author[Spiniello et al.]{C. Spiniello$^{1}$\thanks{E-mail:spini@MPA-Garching.MPG.DE}\\
  $^{1}$ Max-Planck Institute for Astrophysics, 
 Karl-Schwarzschild-Strasse 1, 8l740 Garching, Germany\\}

\author[Spiniello et al.]{\parbox{\textwidth}{C. Spiniello$^{1}$\thanks{E-mail:\texttt{spini@MPA-Garching.MPG.DE}},
N.R. Napolitano$^{2}$,  L. Coccato$^{3}$, V. Pota$^{4}$, A.J. Romanowsky$^{4, 5}$, \\C. Tortora$^{2}$, G. Covone$^{6}$, M. Capaccioli$^{2,3}$
}\vspace{0.4cm}\\
\parbox{\textwidth}{
$^{1}$ Max-Planck Institute for Astrophysics,  Karl-Schwarzschild-Strasse 1, 8l740 Garching, Germany\\
 $^{2}$ INAF - Osservatorio Astronomico di Capodimonte, Salita Moiariello, 16, 80131 Napoli, Italy\\
$^{3}$ European Southern Observatory, Karl-Schwarzschild-Strasse 2, D-85748 Garching, Germany\\
$^{4}$ University of California Observatories, 1156 High Street, Santa Cruz, CA 95064, USA\\
$^{5}$ Department of Physics and Astronomy, San Jos\'e State University, One Washington Square, San Jose, CA 95192, USA\\
$^{6}$ Dipartimento di Scienze Fisiche, Universit\`a di Napoli Federico II, Compl. Univ. Monte S. Angelo, 80126 Napoli, Italy
}
}

\begin{document}

\maketitle
\label{firstpage}


%
\begin{abstract}
We present an integral field study of the internal structure, kinematics and stellar population of the almost edge-on, 
intermediate luminosity ($L_ {*}$) elliptical galaxy NGC~4697. 
We build extended 2-dimensional (2D) maps of the stellar kinematics and line-strengths of the galaxy up to 
$\sim 0.7 $ effective radii (\Reff) using a mosaic of 8 VIMOS (VIsible Multi-Objects Spectrograph on the VLT) 
integral-field unit pointings. 
We find clear evidence for a rotation-supported structure along the major 
axis from the 2D kinematical maps, confirming the previous classification of this system as a `fast-rotator'. 
We study the correlations between the third and fourth Gauss-Hermite moments 
of the line-of-sight velocity distribution (LOSVD) $h_3$ and $h_4$
with the rotation parameter ($V/\sigma$), and compare our findings to hydrodynamical simulations. 
We find remarkable similarities to predictions from gas-rich mergers.  
Based on photometry, we perform a bulge/disk decomposition and study 
the stellar population properties of the two components. 
The bulge and the disk show different stellar populations, with the stars in the bulge 
being older ( age$_{\rm bulge}=13.5^{+1.4}_{-1.4}$ Gyr,  age$_{\rm disk}=10.5^{+1.6}_{-2.0}$Gyr)  
and more metal-poor ( $\mathrm{[M/H]_{bulge}} = -0.17^{+0.12}_{-0.1}$, $\mathrm{[M/H]_{disk}} = -0.03^{+0.02}_{-0.1}$).  
The evidence of a later-formed, more metal-rich disk embedded in an older, more metal-poor bulge, 
together with the LOSVD structure, supports a mass assembly scenario dominated by gas-rich minor mergers 
and possibly with a late gas-rich major merger that left a previously rapidly rotating system unchanged. 
The bulge and the disk do not show signs of different stellar Initial Mass Function slopes,  and 
both match well with a Milky Way-like IMF. 
\end{abstract}

%

%
\section{Introduction}
\label{sect:intro}
Integral Field Spectroscopy (IFS) makes it possible to derive spatially resolved and continuously sampled 
line-of-sight velocity distribution (LOSVD) 2-dimensional (2D) maps, providing a unique view of 
the internal structure of galaxies, as well as detailed morphological and dynamical information, which are essential to 
fully understand the physical mechanisms that regulate galaxy formation and evolution. 

An increasing number of studies have recently derived the integrated and spatially resolved properties 
of a statistical sample of nearby  galaxies  using IFS (e.g.  \citealt{Bacon2001,Thatte2006,Rawle2008,Czoske2008,Scott2012,Fogarty2014,Gonzalez2014,Sanchez2014,Walcher2014}). 
Among the most successful, the \atlas (\citealt{Cappellari2011}) data set provides a complete inventory 
of the baryon budget and a detailed 2D analysis of stellar and gaseous kinematics, together with resolved stellar populations 
of a complete and statistically significant sample of early-type galaxies (ETGs) in the local Universe.

One of the most interesting results of the \atlas\, team is that it is possible to connect 
the 2D features of the line-of-sight-velocity distribution (LOSVD) to galaxy formation scenarios 
by comparing these to high-resolution simulations (e.g. \citealt{Naab2014}, hereafter N+14). 
For the majority of early-type galaxies ($\sim 86$\%), the formation processes result in disk-like objects 
that maintain the axisymmetric shape of the progenitors, while only a small fraction of systems 
(the most massive ones, $12$\%) rotate slowly with no indication of disks. 
Possible processes that can form embedded disk components include minor mergers,  
major merger with disk regrowth, gas accretion events and secular evolution (\citealt{Krajnovic2011}). 

It is possible, in principle, to discriminate between the different formation scenarios 
by accurately analyzing the higher moments of the LOSVD, the Gauss-Hermite coefficients  
$h_{3}$ and $h_{4}$ (representing skewness and kurtosis, respectively) and comparing these to 
high-resolution simulations. 
Gas has a strong impact on the non-Gaussian shapes of the line-of-sight velocity 
distributions (LOSVDs) of the merger remnants. 
In particular, when the values of $h_{3}$ and $V_{\mathrm{rot}}$ have 
opposite signs, the leading (pro-grade) wing is steeper than the trailing (retrograde) wing. This 
indicates the presence of an embedded disk (see e.g. \citealt{Bender1994}).
When an anti-correlation between the $h_{3}$ and the $V/\sigma$ values does not exist but the galaxy 
shows a fast rotation profile, the rotation could be have caused by a recent gas poor major merger 
that typically does not permit re-growth of a significant disk component (\citealt{Barnes1996, Naab2006,Hoffman2009}, hereafter H+09). 
The sign of the $h_{4}$ coefficient distinguishes a peaked shape ($h_{4}>0$) from a flat-topped one ($h_{4}<0$), where the peak
is broad and the wings are narrow.

The age and metal content of galaxies can provide additional important information 
about the formation mechanism and its time-scale. 
To obtain an unbiased and detailed understanding of the star formation 
history of ETGs from their integrated light, it is necessary to study carefully their unresolved stellar 
populations, and to break the degeneracies existing between age, metallicity and element abundances (e.g. \citealt{Worthey1994}).  
The best way to achieve this is to look at as many stellar features as possible (\citealt{Spiniello2014}, hereafter S+14).  
Clearly, in this context, the very narrow wavelength coverage of SAURON is not ideal, since stellar population studies would 
rely on only a few well known blue-optical absorption lines (\Hbeta, \Mgb, Fe5015 and Fe5270).  

The VIMOS IFU (VIsible Multi-Objects Spectrograph on the VLT) covers a wider wavelength range 
(5250-7400 \AA\, using the blue grism and 6450-8600 \AA\, using the red grism, in its high-resolution mode), 
with comparable spectral and spatial resolution, and is clearly a better choice. 
This point is particularly relevant after the recent finding that the low-mass end of the Initial Mass Function (IMF) 
is not universal and might become steeper with increasing stellar mass and stellar velocity dispersion
(\citealt{vandokkum2010, Treu2010, Spiniello2012, Cappellari2012, Cappellari2013, Dutton2013, Spiniello2014, 
LaBarbera2013, Tortora2013,Tortora2014a, Tortora2014b}, but see also \citealt{Smith2015}) or with changes in 
other galaxy structural parameters (e.g. compactness, stellar density, size, see Spiniello et al. 2015b, submitted). 
Indeed, to constrain the low-mass end of the IMF, it is crucial to look at the red part of the optical spectra, 
where most of the gravity-sensitive absorption features are. Some of the most promising indices, like the optical TiO molecular lines  
(bTiO, aTiO, TiO1; S+14), can be captured with VIMOS but are outside the SAURON coverage. 
Moreover, the SWELLS survey (\citealt{Treu2011Swells}) has shown that there could be one universal
normalization of the IMF for disk-like stellar populations (presumably Chabrier-like) and a “heavier” (Salpeter-like) one for the
older and possibly more metal rich stellar populations found in massive bulges and spheroids (\citealt{Dutton2013Swells}). 

Furthermore, 2D integral-field spectroscopy is crucial for constraining the low-mass end of the IMF slope, given the 
very recent observational evidence that seems to indicate that the IMF normalization varies
within individual galaxies (\citealt{MartinNavarro2014}; Sarzi, Spiniello \& Krajnov\'ic, 2015, in prep.) 
following the velocity dispersion profile (but see also \citealt{Smith2015, Spiniello2015b}). 
In this context, a strong observational limitation is the small field-of-view (FOV) of IFS instruments, 
typically of the order of tens of arcsec, preventing good coverage of the target galaxies on the sky and a proper spatially-resolved 
study of the IMF slope.  
Only very recently has a new class of integral-field spectrographs with larger field of view been built to achieve the challenging 
task of studying the internal structure of galaxies. 
Unfortunately most of them still cover a small range of wavelength and therefore are not suitable for constraining the low-mass end of the IMF 
normalization (e.g. VIRUS-P, 3600--6800 \AA,  \citealt{Hill2008}, or CALIFA,  3750--7000 \AA, \citealt{Sanchez2012} in its red configuration).
 The VLT Multi Unit Spectroscopic Explorer (MUSE) represents the perfect solution, combining a large field of view ($1\times1$ arcmin) 
 and a wide wavelength range (4650--9300 \AA). 
Finally, in the  near future the MaNGA (Mapping Nearby Galaxies at APO, \citealt{Law2014}) project will further improve the situation, 
covering the wavelength range 3600--10000 \AA\, up to $\sim 1.5$\Reff\, with a spatial resolution of $\sim 2.5$ arcsec (and a resolution of R$\sim 2000$).
We refer to Table 1 of \citet{Brodie2014} for a detailed comparison between current and future integral-field surveys. 

Thus, at the present day, VIMOS remains among the most competitive instruments for galaxy structure and 
stellar population studies, provided that the small FOV is compensated for by mosaic observations, i.e. contiguous 
IFU pointings astrometrized and stacked together to create a larger final FOV.

This is indeed the approach we decided to follow in order to attempt 
to systematically study the stellar population and the kinematics of the whole surface of nearby ETGs up to almost 
the effective radius over a large wavelength range and with medium-high spectral resolution. 
Given the challenges in the data reduction, calibration and visualization, 
with this paper we present the analysis of one of the galaxies observed in a pilot program focusing on the kinematics 
and the stellar population analysis of two flattened galaxies (NGC~821 and NGC~4697)\footnote{
Program IDs P079.B-0622A and P085.B-0949 (P.I. N.R. Napolitano).}. 
In particular we focus on NGC~4697, for which we have 8 VIMOS pointings
covering  $\sim80''\times60''$ effective area, which allows us to 
investigate regions up to $\sim 45''$ from the galaxy center, i.e. $\sim$0.7~\Reff\,  (\Reff $=66"$ from \citealt{deLorenzi2008}). 

\subsection{Target selection and outline of the paper}
\label{sect:target}
NGC~4697 is an E6, almost edge-on galaxy located along the Virgo southern extension,  
with evidence of substructures (from {\it HST}  ACS images, \citealt{Lauer1995}) and the presence of a 
strong streaming motion component (known since the pioneering work of \citealt{Bertola1975}).  
The galaxy is representative of the class of disky, fast-rotating elliptical $L_ {*}$ 
galaxies (\citealt{Bender1994}), which are candidates for having less dark matter  
(\citealt{Covone2004,Napolitano2005,Cappellari2006}) than the boxy, slow-rotators of comparable mass.
As such, it is among the sample of elliptical galaxies whose planetary nebula (PN) kinematics show 
surprisingly little sign of a massive dark matter halo (\citealt{Mendez2001,Romanowsky2003,Napolitano2005, Douglas2007,Mendez2009}). 
Finally, the spheroidal part of NGC~4697 may well be a large bulge (\citealt{Dejonghe1996}) and therefore
this system is an interesting candidate for testing the SWELLS (\citealt{Treu2011Swells}) claim about the different 
normalization of the low-mass end of the IMF slope (universal, Milky Way-like IMF for the disks and more bottom-heavy IMF, 
possibly steepening with velocity dispersion for bulges, \citealt{Dutton2013Swells}). 

Based on the combined information of the 2D kinematical characterization of the system, 
and stellar population analysis, we are able to fully characterize the stellar population of the disk and bulge components of NGC~4697. 
This is important {\sl per se} in order to investigate the variation of the stellar populations and IMF slope in different structural components 
of the system and possibly to infer a formation scenario for it, but it will also allow us to correctly normalize the stellar 
mass of the galaxy in forthcoming dynamical analysis.

The paper is organized as follows: in Section \ref{sect:observ} we outline the observations and data reduction, in Section \ref{sect:kin} 
we introduce our kinematical technique and we show 2D kinematical maps as well as 1-dimensional (1D) 
kinematics profiles along the major axis of the galaxy. We compare the latter with kinematics profiles in literature.  
In particular, in this Section, we highlight correlations between the $h_{3}$ and $h_{4}$ parameters and  
$V/\sigma$ and try to connect these to  predictions from galaxy formation simulations. 
In Section 4, we focus on the stellar population analysis performed 
via line-index strengths  measurements  on the 1D spectra of the disk and of the bulge separately. 
We also attempt to constrain the IMF normalization of the two components. 
Finally, discussion and general findings are presented in Section 5 together with conclusions.  

\placefigfov

\section{Observation and data reduction}
\label{sect:observ}
The integral-field spectroscopic (IFS) observations have been carried out with VIMOS, 
the multi-mode wide-field optical instrument mounted on the ESO VLT, 
as part of a proposal carried on in two different runs : May 2007 and March 2010, respectively. 

The Integral-field Unit (IFU) head of VIMOS consists of a square array of 
$80 \times 80$ micro-lenses with a sampling  of $0.67\arcsec$ per spatial element 
(‘spaxel’) at the Nasmyth focus of the telescope. 
In its high-resolution mode only the central 1600 lenses are used, resulting in a field-of-view (FOV) of $27" \times 27"$. 
The seeing was generally below $0.8\arcsec$ during the observations, so that each fiber of the IFU gives a spectrum 
that is essentially independent of its neighbors.
All data sets have been taken with the high-resolution 
blue grism, covering a wavelength range between $4100 - 6300$ \AA\, 
with a dispersion of $0.51$ \AA/pixel.
The spectral resolution of the HR-Blue grism is $\lambda / \Delta \lambda = 2550$, where 
$\Delta \lambda$ is the full width at half-maximum (FWHM) of the instrument profile. 

In order to cover the effective radius and to have a high S/N among the whole FOV,  we use a mosaic approach. 
During the first observation run, we have obtained 6 different pointings,  
each made by six exposures of $\sim 800$ seconds, while for the second run we 
have obtained 2 pointings along the major axis of the system,  with 4 exposures of $825$ seconds each.
The Position Angle of all pointings has been fixed to $PA=-20°$.
For accurate measurements of the stellar kinematics, one must properly evaluate the sky background 
contamination. Since VIMOS does not have sky-dedicated fibers, and the target galaxy covers the entire 
field-of-view, it has been necessary to make separate telescope pointings (of $\sim 300$ seconds) to measure 
the sky background for each observing night.

\subsection{Data Reduction}
\label{subsect:datared}
Data reduction has been carried out using the VIMOS pipeline (version 2.1.1) by the ESO  
``Data Flow System Group\footnote{N. Devillard, Y. Jung, R. Palsa, C. Izzo, P. Ballester, C. Sabet, K. Banse, M. Kiesgen, L. Lundin, 
A. Modigliani, D. J. McKay, N. P. F. McKay, J. Moeller-Larsen, L. de Bilbao}", 
via the Gasgano data file organizer developed by ESO. 
A schematic representation of the data reduction processes performed within Gasgano is provided by \citet{Zanichelli2005}. 
The reduction involves standard steps for fiber-based integral-field spectroscopy:  
\textit{(i)} Subtraction of a master bias frame created by averaging all the bias frames  
\textit{(ii)} Fiber identification (correct association of a fiber position on the IFU mask to a corresponding position 
      on the CCD) and correction of the differences in the fiber-to-fiber transmission
\textit{(iii)} Wavelength calibration 
\textit{(iv)} Scientific image combination.

We create a final data cube for each pointing by averaging over the final single-observing block (OB) 
flux-calibrated science cubes produced by the pipeline 
using the IFU tables, with the correspondence between each row (i.e. each spectrum) 
of the extracted spectra and the fiber's position on the IFU head. 
The four IFU quadrants  are treated independently up to the combination of all exposures  into a single-pointing datacube.

We subtract the median sky spectrum (extracted directly from the `sky frame') from the `science frame'. 
Median combining was needed to ensure that any residual contamination from faint objects was removed. 
We are confident that this sky-subtraction procedure is accurate enough but, if prominent sky emission lines are 
still present in the spectra after sky subtraction, we mask them out during the the stellar kinematics recovery process. 
We also investigate possible effects of the differential atmospheric refraction by checking whether or not the  
photometric center of the galaxy changes its position as a function of wavelength in the datacube. 
We find that the position does not change by more than 2 pixels ($\sim 1.3$ arcsec) between 4000 and 6200 \AA. 
Each single-pointing datacube is astrometrically calibrated independently by matching the position of the 
photometric center or of a star (when the center is not visible in that particular pointing) 
to an {\it HST} ACS image of the central region of NGC~4697. Subsequently,  the different pointings are  
linearly interpolated onto a common 3D grid uniformly sampled,  after normalizing them for exposure times, 
using a simple grid interpolation \footnote{We interpolate, shift and rotate each single-pointing datacube 
by matching the position of the center and the rotational angle into this datacube-container.} to obtain the final mosaic datacube. 
As a further check of the robustness of this mosaic reconstruction, 
we re-check the WCS astrometry on our final combined image, 
using as reference the same {\it HST}/ACS image. 

The results of this comparison are visible in 
 Figure~\ref{fig:fov} which shows the final reconstructed mosaic image, obtained by averaging the datacube along 
the wavelength direction, as well as the ACS image.  Twenty blue isophotes, derived from the {\it HST} image, 
 are shown on both images to check the accuracy of our astrometrical calibration. Although most of the point sources visible in 
 the {\it HST} image are too faint to be detected in our VIMOS observations, the positions of the center and of a star (ENE) 
 coincide between the two instruments.
On the right side of Figure~\ref{fig:fov} we also plot a 1D spectrum extracted at the photometric
center of the galaxy and another 1D spectrum extracted at $\sim 40"$, in a position which is highlighted in the reconstructed FOV 
by a small yellow star.


\section{Kinematics Measurements}
 \label{sect:kin} 
To recover the line-of-sight velocity distribution (LOSVD) first, 
 we apply a spatial binning method in order to increase the signal-to-noise ratio (S/N) 
over the single-fiber spatially resolved spectra of the final datacube (Section 3.1). 
Then we measure the LOSVD for the  
galaxy using a maximum penalized likelihood approach (Section 3.2).  
In this way, we obtain spatially resolved 2D profiles of rotation velocity, V, and  velocity dispersion, $\sigma$, 
as well as the higher order Gauss-Hermite moments $h_{3}$ and $h_{4}$, 
which quantify the asymmetric and symmetric departures of the LOSVD from a pure 
Gaussian (related to the skewness and kurtosis respectively).
In the following sections we describe in more detail the spatial binning procedure and the LOSVD recovery.

\subsection{Voronoi 2D spatial binning}
\label{subsect:voronoi}
We re-bin the data with a fixed final S/N by locally averaging
neighboring fibers (using the Voronoi Adaptive technique IDL code of \citealt{Cappellari2003}). 
We use an adaptive binning scheme where the size of the bin is adapted to the 
local S/N. Bigger spatial bins are created in the low-S/N regions, 
with higher resolution and smaller bins retained in the high-S/N 
regions (towards the galaxy center)\footnote{Spaxels are arranged in a regular grid, so the size is a well defined quantity.}.

We tested several values for the final Voronoi fixed S/N, choosing a final threshold of  $S/N=70$. 
In this way we are able to obtain a high resolution spectrum for each spaxel, which is fundamental for performing a proper 
stellar population analysis based on line-index measurements, while also preserving spatial information 
that is crucial for recovering detailed spatially resolved kinematics information up to $\sim 0.7$ effective radius (\Reff$=66"$). 

\subsection{Line-Of-Sight Velocity Distribution (LOSVD) fitting}
\label{subsect:losvd}
We measure rotation velocity, velocity dispersion and higher-order moments (V, $\sigma$, $h_{3}$, $h_{4}$) 
for each Voronoi-rebinned spaxel spectrum using the pPXF code of \citet{Cappellari2004}. 
This software, working in pixel space, finds the combination of stellar templates which, convolved 
with an appropriate line-of-sight-velocity distribution, best reproduces the galaxy spectrum.
The best-fitting parameters of the LOSVD are determined by minimizing $\chi^{2}$, which 
measures the agreement between the model and the observed galaxy spectrum, over the set of N good pixels.
We use the MILES Library of Stellar Spectra (\citealt{SanchezBlazquez2006, Falcon-Barroso2011}): 
985 well-calibrated stars covering the region from 3525-7500 \AA\,  at a spectral resolution of 2.54 \AA\, (FWHM),  
$\sigma \sim 57$ \kms\, at the central wavelength, obtained at the 2.5m Isaac Newton Telescope (INT). 
The pPXF code also allows the user to mask out noisy regions of the galaxy spectrum. 
We mask out sky emission lines, focusing on absorption lines from $4300-5400$ \AA\, 
(including in this way \Hbeta, O lines, \Mgb\, doublet and Fe). 
Uncertainties on individual points are obtained via a Monte Carlo simulation.  

We follow the approach described in \citet{Cappellari2004},  to recover rotation, velocity dispersion and the higher-order terms 
of the Gauss-Hermite series simultaneously.  
The method makes use of a penalization factor to the solution, which biases the results to suppress the noise. 
Obviously, one would like the penalty to leave the LOSVD virtually unaffected. 
Thus, we first  perform the kinematics fit without including any penalty (i.e. setting the keyword BIAS=0 in the pPXF code) 
to make sure that our results have not been biased and also to properly estimate the uncertainties on the velocity distribution. 
This produces a noisy solution (especially for $\sigma$) but allows us to set a range of values for $h_{3}$ and $h_{4}$
which can be used to estimate the right value for the penalization factor. 

In particular, this has been done with the {\sl `ppxf simulation'} procedure (included in the publicly available code). 
The routine performs a Monte Carlo simulation of the spectra, which repeats
the full fitting process for a large number of different realizations of the data obtained by randomly adding noise to the original
spectra with a resulting S/N equal to the chosen threshold value (S/N = 70 in our case), and using   
as input for [$h_{3}$ , $h_{4}$] the maximum representative values measured in the non-penalized pPXF fit of the previous step ([$-0.3,0.3$]). 
At this point, we repeat the fitting procedure while using as penalty the largest value such that the mean difference between the 
output and the input [$h_{3}$ , $h_{4}$] is well within the rms scatter of the simulated values (BIAS=1.5). 

Finally, we inspect by eye zones of the FOV where we expect to have contamination 
(e.g.\ near the brighter stars) and spectra for which the reduced $\chi^{2}$ of the fit was larger than 3.  
In particular, we identify  $\sim 30$  ``problematic" Voronoi cells, for which we go back to the 
single fiber spectra and re-create the Voronoi cell while excluding 
contaminated or noisy spectra. Then we re-run the pPXF procedure, in some cases also 
setting by hand the pixel regions where the fit is performed. 

We take into account all the main systematic uncertainties, mostly caused by template mismatch, and wavelength coverage. 
In order to minimize template-mismatch errors, we fit a broad set of stellar templates 
together with the kinematics, after convolving both to a common resolution. 
We also test the robustness of the kinematic results by changing the spectral region 
of the fit and we find results always consistent within $2\sigma$  (not including systematics). 
The uncertainties on the final inferred kinematics are estimated by adding in quadrature the formal uncertainty 
given by the Monte Carlo simulation and the scatter in the results for different templates and spectral regions.
Typically, we find uncertainties of the order of 5\% for rotation and velocity dispersion and at most of the order of 10\% 
for the higher-order LOSVD moments.
For the central bin, we obtain a velocity dispersion of $\sigma_{\star}= 161 \pm 8$ \kms\, 
perfectly consistent with previously published results (i.e. \citealt{Binney1990, deLorenzi2008}).

\subsection{Two dimensional kinematics maps}
\label{subsect:2Dmaps}
The reconstructed mosaic ($\sim80" \times 60"$ effective area), is among the largest 2D maps 
ever measured for an elliptical galaxy and provides a powerful tool to full characterize the internal kinematics of the system 
and the spatially-resolved properties of its underlying stellar population, which are necessary ingredients to understand the dynamics 
and its connection with formation mechanisms. 
Here below, we characterize the 2D behavior of all the relevant kinematical quantities, while in the next sections 
we discuss the properties of the stellar populations and try to establish their connection with the kinematical structure. 
The full dynamical analysis as well as the derivation of the mass and orbital profile of the galaxy 
will be presented in a forthcoming paper.

\placefigmappe

Figure~\ref{fig:mappe} shows 2D maps of rotation velocity, velocity dispersion 
and the Gauss-Hermite moments $h_{3}$ and $h_{4}$ of the LOSVD. On the left column, these quantities are
shown as derived from the Voronoi-binning technique discussed in Section \ref{subsect:voronoi}. 
Here the small scale features are visible, mainly in the center, while the large scale structure at larger radii looks noisy. 
In order to make more evident the large scale pattern of the kinematical quantities, we show on the right column the 2D 
quantities smoothed by averaging the Voronoi values over a grid of 15$\times$15 macro bins and then
interpolated. The final average maps have been masked with an ellipse mimicking the orientation and the 
axis ratio of the outermost isophote enclosing the VIMOS mosaic.

The galaxy shows a rotation velocity (upper panel) and $h_{3}$  (third panel from above) patterns typical of a
 rotation-supported structure along the major axis of the system. In particular, the $h_{3}$ indicates the presence of a 
disk-like structure elongated along the major axis. 

The velocity dispersion map (second panel), on the other hand, is rather flat over the full area covered by our observations, 
with minimal, statistically insignificant fluctuations. 

Finally the $h_4$ parameter map (bottom panel) shows a clear elongated 
area along the line of nodes with positive values ($h_{4}\sim 0.01$), which tend to become negative 
in the regions off the plane. 
which are not common for disk-like structure.
The presence of negative $h_{4}$ off the main rotation axis in the regions where the rotation field 
is almost zero indicates tangential orbits, possibly belonging to the bulge component 
that starts to dominate outward (see Sec~\ref{sec:bulgedisk}).
The presence of two components along the line of sight might be responsible for the 
positive $h_{4}$ parameter aligned along the galaxy major axis by producing a 
overall peaked distribution due to the superposition of a cold disk
and hot bulge. 

IFS provides a complete and high resolution 2D map of the LOSVD, 
allowing the definition and estimation of a number of important parameters that can 
be used to better understand the detailed internal structure of the galaxy, a crucial ingredient
for investigating possible formation scenarios. 
An interesting quantity that is possible to derive in 2D is the ratio of the ordered 
and random motion in a galaxy ($V/\sigma$).

\placefigVsigma

In Figure~\ref{fig:vsig} 
we present the first  $V/\sigma$ 2D map of NGC~4697, up to $\sim 0.7$\Reff ,which gives a 
more quantitative view of the dynamical state of the galaxy
and further confirms the presence of the galaxy sub-components. 
One expects that the regions which are bulge-dominated 
have smaller $V/\sigma$ values, while the rotation-supported regions where 
the disk is dominant have larger $V/\sigma$ values. 
This is indeed what we observe in the 2D map and it is further confirmed in the next Section, 
where we compare our $h_{3}$--  and $h_{4}$--$V/\sigma$ correlations with those inferred from 
hydrodynamical simulations. 

We note that none of the previous 3D measurements of galaxies have performed the same kinematical diagnostic as the one above. 
However, more standard rotation and dispersion 2D maps of the LOSVD of NGC~4697, like the ones 
derived from the ATLAS$^{\rm 3D}$  survey, are fully consistent with the ones presented here, 
although  with a lower spatial resolution and over a smaller FOV (ATLAS$^{\rm 3D}$ covers $40\times60$ arcsec, \citealt{Krajnovic2011}).   
We do not directly compare our 2D maps to those obtained by \atlas\ because of the different spatial and spectral resolution but 
in  Section 3.5 we extract 1D profiles along the major axis and compare our kinematics profiles against the \atlas ones. 

Finally, extended 2D kinematical maps up to  $3.5$ \Reff\, of this galaxy were recently presented as part of 
the SLUGGS Survey (\citealt{Arnold2014}).  Overall there is a fair agreement on the 
features observed in the data, both in the velocity field and in the higher-order moments of the LOSVD. 
In fact, based on their more spatially-extended kinematic map, the SLUGGS team concluded that the 
fast rotation is due to an embedded disk with a rotation amplitude clearly declining with radius. 
The presence of a disk  is also found from our detailed analysis of the $h_{3}-$ and $h_{4}$--$V/\sigma$ diagrams.  

\subsection{Comparison with simulations}
$h_{3}$--$V/\sigma$ and $h_{4}$--$V/\sigma$ correlations have also been 
predicted in simulations of disk merging (\citealt{Naab2006b}, H+09 and N+14). 
For instance, N+14 linked the present day central kinematic properties of galaxies 
to their individual cosmological formation histories and showed that galaxy mergers (major and minor) 
have a significant influence on the rotation properties, resulting in both a spin-down and a spin-up 
of the merger remnants, and in a clear anti-correlation between $h_{3}$ and $V/\sigma$.

N+14 have defined six classes (named with letters from A to F) 
with different evolutionary paths and histories of mass assembly that can be 
directly linked to their 2D kinematic maps and the present day shapes. 
The galaxies groupings are based on the 2D kinematic maps and on the 
so-called  `spin' parameter (\lambdaR), a proxy for the projected  
angular momentum per unit mass (\citealt{Emsellem2007, Emsellem2011}).   

The \lambdaR\, parameter has been used to show that, within the effective radius, 
a dichotomy exists between `fast-' (\lambdaR$>0.1$) and `slow-' (\lambdaR$<0.1$) 
rotators (\citealt{Emsellem2007}). 
The second class of galaxies shows little or no rotation, significant 
misalignment between the photometric and the kinematic axes, 
and contains kinematically decoupled components, 
whereas the first class exhibits regular stellar velocity fields, 
consistent with disk-like rotation and sometimes bars. 
This version of the `spin' parameter takes the form:
\begin{equation}
\lambda_{R} = \frac{\sum_{i=1}^{N_{p}} F_{i}R_{i} |V_{i}|}{\sum_{i=1}^{N_{p}} F_{i}R_{i} \sqrt{V^{2}_{i}+\sigma^{2}_{i}}} 
\label{lambdaR}
\end{equation}
where $F_{i}$, $R_{i}$, $V_{i}$ and $\sigma_{i}$ are the flux, circular radius, 
velocity and velocity dispersion of the i-th spatial bin, 
with the sum running over the $N_{p}$ bins. 

A revised and more accurate separation of these two classes of galaxies
has been proposed  by \citet{Emsellem2011} to include the associated apparent ellipticity:
a higher value of the specific stellar angular
momentum is expected for galaxies which are more flattened or closer to edge-on
if these are all intrinsically fast rotators\footnote{A galaxy with a relatively low value 
of \lambdaR, e.g., of 0.2, may be consistent with a simple spheroidal axisymmetric 
system viewed at a high inclination (near face-on), but this is
true only if its ellipticity is correspondingly low (e.g., $\varepsilon \leq 0.3$). 
A large ellipticity would instead imply a more extreme object (in terms of orbital
structure or anisotropy).}. 

The threshold for \lambdaR\, was fixed to be 
proportional to square-root of  $\varepsilon$ with a scaling parameter
 that depends on the aperture (see Appendix B in \citealt{Emsellem2011} for further details). 
For instance, at $R_{\rm eff}/2$

\begin{equation}
\lambda_{R_{\rm eff}/2} = (0.265 \pm 0.01) \times \sqrt(\varepsilon_{eff/2})
\label{lambdaR2}
\end{equation}

Following this classification, NGC~4697 was previously classified as ``fast-rotator''.  
This means that, given an ellipticity of $\varepsilon= 0.36$ (\citealt{Peletier1990}), 
one expects for this parameter  a value of \lambdaR$>0.16$ (for $r<R_{\rm eff}/2$). 
We find $\lambda_{R_{\rm eff}/2}=0.35 \pm 0.05$, confirming the classification of 
this galaxy as a `fast-rotator' and in perfect agreement with the \atlas 
result ($\lambda_{R_{\rm eff}/2}=0.322$, \citealt{Emsellem2011}). 

Fast-rotators belong to class A, B and D in the N+14 classifications. 
However only the first two groups of galaxies will result in $h_{3}$ and $h_{4}$ 
distributions similar to that we observe in NGC~4697. 
Both groups (A and B) of galaxies are rotationally supported, had a significant amount of central in-situ, 
dissipative, star formation but, while class A have late assembly histories dominated by minor mergers, 
class B galaxies have gone through a late gas-rich major merger. 
These two formation scenarios will produce the same $h_{3}$--$V/\sigma$ anti-correlation 
that we observe in our maps. 

One way to distinguish between the two classes is to look at the  \lambdaR\, profiles:
class A galaxies show peaked \lambdaR\, profiles within the effective radius, 
originating from the fast rotating central stellar disk, whereas class B galaxies have \lambdaR\, 
profiles constantly rising beyond the effective radius. 

\placefiglambdar

In Figure~\ref{fig:lambdaR} we plot the  \lambdaR\, profile as a function of radius calculated from circular apertures. 
The profile has a strong peak around $0.2$\Reff\, and then it rises until $\sim$\Reff, which  
suggests that the mass assembly of NGC~4697 has involved significant in-situ star formation, 
 that NGC~4697 has experienced several gas-rich minor mergers 
(like simulated galaxies of class A)  and possibly a late gas-rich major merger 
leaving a previously rapidly rotating system unchanged (like simulated galaxies of class B). 
However given the poor resolution of the simulations this interpretation should be taken with caution (Fig.5 in N+15
shows the \lambdaR\, profiles of the simulated galaxies in the six classes and  only one galaxy of class A shows 
a peak, which is at $\sim 0.95$\Reff. On the other hand we observe 
that in Fig.5 of \citealt{Emsellem2011}, showing  \lambdaR\,
for the full \atlas\, sample,  several fast-rotators show a profile similar to what we derive for NGC~4697). 
Moreover, \citet{Coccato2009} derived the \lambdaR\, profile for NGC~4697 out to $\sim 3.5$\Reff\, based 
on planetary nebulae (PNe) and showed that it flattens out at  $\sim 1$\Reff\, which supports the idea that 
NGC~4697 should be compared with simulated galaxies of class A. 
We also note that \citet{Arnold2014} concluded that galaxies like NGC~4697 formed through 
rapid, dissipative processes, which formed the central regions, followed by a prolonged period of dry minor mergers.
We cannot exclude the scenario of gas-rich late major mergers, but independent lines of evidence point toward
a scenario where the mass assembly of NGC~4697 is dominated by minor mergers and gradual dissipation.

\placefigsimul

In Figure~\ref{fig:simulh3h4} we show the $h_{3}$--  and $h_{4}$--$V/\sigma$ 
correlations derived from the 2D kinematic maps (left panels), and we compare them to 
luminosity contours derived by H+09 for a 1:1 disk--disk merger with 30\% gas fraction (middle panels) 
and from  the cosmological hydrodynamical simulations of N+14 (class A and class B, right panels). 
 A  good agreement, at least on a qualitative level, is found between data and simulations, 
although we caution that H+09 only considered the case of equal-mass mergers.

VIMOS data-points are color coded in accord with their value of $|Cos[\theta]|$, 
calculated with respect to the galaxy minor kinematical axis.
We find a clear correlation between the $|Cos[\theta]|$ values and the regions where 
the disk component is elongated. 
Points oriented near the major axis (i.e., $|Cos[\theta]| \sim 0$) have also 
higher absolute values of $h_{3}$ and therefore are associated with the rotation-supported structure. 
Similarly, points with higher values of $|Cos[\theta]|$ have lower $h_{3}$ and roughly 
correspond to the bulge region. 
The $h_{4}$ does not show a similar (anti-)correlation with $V/\sigma$.  

The $h_{3}$--$V/\sigma$ anti-correlation generically tracse cold disks. 
It is a feature that naturally emerges in 
 disk-disk gas-rich mergers, and tends to produce large asymmetric deviations 
 from Gaussianity because of the presence of sub-components that retain a memory 
of the disk-star streaming, and because of gas that settles down through dissipation to form a thin embedded disk.
This $h_{3}$--$V/\sigma$  correlation is generally negative when the dominant orbit population tends to stream 
(e.g. z-tube orbits, see also \citealt{Rottgers2014}), and positive when the rotation is caused by sub-dominant 
streaming population (e.g. x-tube orbits) 
superimposed on the tail of the dominant streaming orbits (e.g z-tubes which still contribute  some rotation along
the minor axis; see also the discussion in H+09). 
The same features are not found in second generations of major (re-)mergers 
(mergers at lower redshift, where less cold gas is present and with elliptical-morphology progenitors), in which case 
 the features are more mixed and look more concentrated around the center of the diagram. 

The $h_{4}$--$V/\sigma$ correlation of NGC~4697 in Figure~\ref{fig:simulh3h4} shows a V-shaped distribution
with a double peak at $|V/\sigma|\sim0.5$.
This distribution is very similar to  the predictions from simulations of N+14 (right, bottom panel)  
and it also agree with H+09 (middle, bottom panel), although there is a significant 
offset between the data and the H+09 predictions. 
The peaked population could correspond to a late-formed disk stars (note the correspondence to the extreme
$h_{3}$ values and $|Cos[\theta]| \sim 0$ indicating the alignment along the major axis) which tend to produce a positive 
$h_{4}$ in superposition with the more isotropic component (differences in ages of the two components 
are discussed in Sec.~\ref{sect:stellpop}). 
In the simulations of H+09 the re-mergers would look more concentrated around the origin and would 
tend to have LOSVDs with positive $h_{4}$, similarly to expectations from violent relaxation, 
which generally results in a radially-biased distribution function (\citealt{Binney2008}). \\
\indent To conclude,  the high-order moments of the LOSVD provide kinematical evidence 
for a two-component system: 
the largest absolute values of $h_{3}$ and $h_{4}$ 
indicate an area of the galaxy with the dominance of x-tube over z-tube orbits, which are produced in
collisionless mergers and trace the presence of a (later) streaming structure,
these are also the regions around $|Cos[\theta]| \sim 0$ (minor axis). 
Values with $h_3\sim 0$ and $h_4\sim 0$ indicate instead the presence of a non rotating subsystem 
that dominates in the regions where the values of $|Cos[\theta]|$ are higher.  
The $h_{3}$--$V/\sigma$ and the peaked \lambdaR\, profile suggest that gas-rich minor mergers 
dominate the mass assembly of NGC~4697, although we cannot completely exclude the possibility of a 
major merger event that happened at redshift $z \geq 1$. 

We make use of this evidence in Section \ref{sec:bulgedisk} where, based on the photometry 
of NGC~4697, we separate the disk from the bulge and study the ages and metal 
abundances of the stellar populations of these two components. 

The details of the gas accretion leading to the formation of the central disk may be somewhat 
different than the one investigated in the simulation of disk merging and it might be useful to have
LOSVD predictions from simulations of cold accretions as an alternative high dissipative process forming the disk. 
It seems evident, though, from H+09 and N+14 that dry mergers is not able 
to produce the  kinematical substructures we see in the $h3$-- $h4$--$V/\sigma$ diagrams. 
In fact, in a merger without a dissipative gas component, the remnant galaxies are axisymmetric and
do not show a population of stars with high angular momentum tube orbits nor the re-growth 
of a significant disk component (e.g.\ class D in N+14). 

\subsection{One dimensional spatially resolved kinematics profiles}
\label{subsect:onedprof}

\placefigoneDprof

To compare our results against published results based on long-slit data 
and IFS with different FOV and resolution, we extract from the 2D kinematics maps the information along the major axis. 
Specifically, we extract and sum all the spectra positioned in a rectangle with width of $1.34''$ (2 pixels) oriented along 
the major axis of the galaxy and centered on our kinematic center (to be as close as possible to the configuration of slit-based studies).  
We perform the same extraction for the IFU \atlas observations (publicly available on the Survey Webpage) 
in order to perform a simple comparison of the two sets of data with different spatial and spectral sampling.  
In Figure~\ref{fig:delor} we compare our profiles (black points) with the \atlas ones (green points) and with the ones published in 
de Lorenzi et al. (2008, hereafter DL08), derived from high S/N integrated absorption line spectra obtained by \citet{Mendez2005} with 
FORS2 at the VLT, a slit width of 1 arcsec and seeing 1 -- 1.5 arcsec (blue points). 
For the major axis inclination, we assume the value PA = 63 degrees from \citet{Binney1990}, slightly different 
from the one assumed by DL08 but providing the best symmetry and maximum rotation for our extracted 1D profiles. 
We do not worry about this small PA discrepancy that could be caused by the fact that the slit in DL08 was not 
exactly on-center or could be related to the offset we found between kinematic and photometric center, most 
probably caused by a small effect of  atmospheric refraction. 
For the \atlas\, data we use the same inclination of PA = 63 degrees and extract the major 
axis profiles with the same procedure applied to our VIMOS data. 
We found an overall good agreement between the three 1D kinematical profiles which 
 nicely demonstrates the robustness of our mosaic approach and the validity of our results.


\section{2D Stellar Population Analysis}
\label{sect:stellpop}
In this section we study the stellar population of the disk and of the bulge of NGC~4697 separately, 
to put constraints on the age, \alphafe, metallicity and IMF normalization 
of the stellar population of the different morphological components of the galaxy.

To constrain the stellar population parameters, we focus on line-index measurements rather than full
spectral fitting, which avoids issues with spectral calibration when
comparing to observations that might have been poorly calibrated and
we assume a simple stellar population (SSP), rather than an extended star formation history,  
which is still a strong assumption and will be properly addressed in future work.  

We measure equivalent widths (EWs) of several stellar absorption features in each of the Voronoi-rebinned 1D 
spectra, after convolving them to a common resolution of $350$ \kms\, to correct for kinematic 
broadening\footnote{This final resolution has been chosen for practical reasons:
the version of the simple stellar population models used in this paper was created and used in S+14
with a resolution of  $350$ \kms. To save computer power, we did not perform again the EW measurements 
and simply smoothed our data to the resolution of the models. We are confident that this did not cause any 
loss of information, and we check (by eye-inspection) that contamination from sky line residuals do not affect the spectral features.}. 
We compared these EWs to those measured from an extended version of the SSP models of  Conroy \& van Dokkum 
(2012a, hereafter CvD12) with the same common resolution, obtained in Spiniello, Trager \& Koopmans (2014b). 

CvD12  optimized models over the wavelength interval $0.35\, \mu m \,< \lambda < 2.4\,\mu m$ 
at a resolving power of R $\sim 2000$, specifically for the purpose 
of studying old, metal-rich stellar populations.  The models are built combining two 
empirical stellar libraries (MILES, \citealt{SanchezBlazquez2006}; and IRTF, \citealt{Cushing2005})
and three sets of isochrones to measure the IMF slope down 
to $\sim 0.1\, {M}_{\odot}$. The CvD12 models explore variations in age in the range 3--13.5 Gyr,
$\alpha-$enhancement of 0--0.4 dex, and four different single-slope IMFs: 
a bottom-light single-slope power law with $x=1.8$ (where $x$ is the IMF slope, 
using $dN/dm \propto m^{-x}$) that produces a stellar mass-to-light ratio comparable to the one 
obtained from a standard Chabrier IMF (\citealt{Chabrier2003}), a Salpeter (1955) IMF 
with a slope of $x=2.35$, and two bottom-heavy (dwarf-rich) IMFs with slopes of $x=3.0$ and $x=3.5$.

The CvD12 models also allow for variations in the $[\alpha/$Fe] ratio as well as variations in 
the abundance pattern of 11 different single elements.  This is enough to decouple 
IMF variations from abundance variations. 
However, all of the CvD12 models use {\sl solar} metallicity isochrones, even when synthesizing
with different abundance patterns or different $[\alpha/$Fe]. 
The abundance variations of single elements are implemented at fixed [Fe/H], which implies  that the total metallicity 
$Z$ cannot be easily controlled.  This is restrictive, if one aims to disentangle elemental enhancements, 
metallicity changes and IMF variations, especially in the case of  massive ETGs, which have star formation histories 
different from the solar neighborhood (e.g., \citealt{Peterson1976,Peletier1989,Worthey1992, Trager2000b, Arrigoni2010}).  

To resolve this issue, we follow the approach presented in \citet{Spiniello2015}, hereafter S+15 
of extending the parameter space of the SSP models using response functions. 
We construct a metallicity-response function for each given age and IMF slope 
taking the ratio between two spectra of the MIUSCAT SSP models (\citealt{Vazdekis2012})\footnote{The MIUSCAT SSP has 
instead  a fixed total metallicity and covers six metallicity bins 
$Z=0.0004, 0.001, 0.004,0.008,0.019$ and $0.03$, where $0.019$ represents the solar value.} 
with the same age and IMF slope but different total $Z$. 
We then multiply this response function with the spectrum of a CvD12 model with fixed age and IMF 
to build a new model (SSP) that extrapolates the latter model to a new part of parameter space 
(i.e. covering super-solar $Z$). 
These modified SSP models combine the flexibility  of both the MIUSCAT and CvD12 models  
to predict spectra in a part of parameter space that neither of them reaches separately 
(MIUSCAT models use different total metallicities but do not allow one to change the
abundances which are fixed to solar). 
As pointed out in S+14 and S+15, selecting IMF-dependent features that are age- and metallicity-independent 
and combining them with indices that depend mainly on age or mainly on element abundance 
is important for breaking the age-metallicity-IMF degeneracy when using 
SSP models to infer the stellar populations from unresolved galaxies spectra.

We measure some of the classical blue Lick indices (e.g. \Hbeta, \Mgb, Fe5270, Fe5335, [MgFe]: 
\citealt{Gonzalez1993, Worthey1994, Trager2000}) that have been extensively and successfully used in the literature to 
constrain the age and metallicity of galaxies, together with some of the bluer 
IMF-sensitive features defined and used in S+14 and S+15. 
Unfortunately  this paper makes use of data taken with the HR-blue grism, and therefore only covers 
the bluer IMF-features of S+14 (specifically bTiO, aTiO, TiO1) 
while in the future we plan to obtain new data using the HR-red grism which will cover also 
TiO2, CaH1, TiO3 and TiO4, allowing a more detailed analysis and a more secure inference on the low-mass end of the IMF slope. 

Before measuring EWs, we convolved galaxy and model spectra to an effective velocity dispersion of
$\sigma = 350\,\mathrm{km}\,\mathrm{s}^{-1}$ to correct for kinematic broadening. 
Indices in both the galaxy and the model spectra have been measured
with the same definitions and methods\footnote{We use the code SPINDEX from \citet{Trager2008}.}. 
We do not place our indices on the zero-point system of the Lick indices and quote them
as index strengths in units of angstrom (if atomic) or magnitudes (if molecular, like the TiO lines). 

\placefigelemmaps

The great advantage of integral field spectroscopy is the simultaneous availability of spatial and spectral information. 
This allows us to generate 2D spatial maps of spectral features of interest, like the one in Figure~\ref{fig:element_maps}, 
showing  the [MgFe]\footnote{[MgFe] = $\sqrt{(\mathrm{Fe5270} +\mathrm{Fe5335})/2 \times \mathrm{Mg b}}$, (\citealt{Gonzalez1993})} 
index that has been widely used to constrain the metallicity. 
The 2D map suggests different stellar populations in the bulge and in the disk: specifically,  
the [MgFe] EWs are systematically higher in the Voronoi cells within the disk. 
Therefore, the 2D element maps indicate that the stars in the disk are more 
metal-rich than the stars in the bulge. However, we caution the reader that it is not possible to directly infer
the total metallicity by looking at a single-element map, because of the degeneracies present between the stellar 
population parameters. For instance, in this case the [MgFe] values could be higher due to an increase of age and/or could 
increase of total metallicity. This is an example of the so-called age-metallicity degeneracy which can be broken by 
using more than one indicator.  
This motivates us to decompose the bulge and disk components and to extract a 1D spectrum 
of each of them in order to study their stellar populations separately through a  more quantitative, 
proper SSP analysis.  

\placefigoneDspec

\subsection{Disk-Bulge decomposition}
\label{sec:bulgedisk}
We perform a two-component fit to the HST V-band image of NGC~4697 obtained by \citet{Tal2009}, driven by the 
correlations in the $h_{3}$ and $h_{4}$ parameters versus $V/\sigma$ that clearly indicate the presence of a disk. 
We use the latest version of GALFIT (\citealt{Peng2010}) to fit two ellipsoid models (`sersic' for the bulge and `expdisk' for the disk) 
to the light profiles of the galaxy image. 
We obtain a similar position angle (PA $\sim 64$ deg, confirming previous findings) 
but very different axis-ratios for the two components: $(b/a)_{disk}=0.297$,  $(b/a)_{bulge}=0.7512$.   
We use those to spectroscopically separate the bulge from the disk. 
We note that a more rigorous spectroscopic separation, following the approach 
of  \citet{Coccato2011} and \citet{Coccato2014}, will be performed in future studies to fully separate these two components.  

First, we extract and sum together all the spectra in the datacube with $r < $\Reff$/8$. 
In this central region we expect bulge and disk to both contribute substantially. 
Second, we select and sum the spectra belonging to an elliptical anulus centered 
on the kinematics center with the axis-ratio and PA found by GALFIT for the disk: $(b/a)_{disk}=0.297$, $PA=65.47$deg 
and with $r > $\Reff$/8$ (to exclude the central region). 
Finally, we sum up the remaining spectra where the contribution of the bulge is dominant. 
The resulting 1D spectra have very high S/N ($S/N = [450, 320,280]$ for center, disk and bulge respectively). 

In Figure~\ref{fig:1dspectra} we show the extracted 1D spectra of the disk and
bulge, as well as of the mixed (disk$+$bulge) population of the central region of NGC~4697 for comparison. 
Shaded yellow bands show some of the line indices used in this work, along with continuum side-bands (blue and red colored bands). 
Some small differences are visible already from the spectra: the disk spectrum shows stronger Mg and Fe lines, 
which confirms what have already seen from the [MgFe] 2D map and gives hints of a more metal-rich population. 
The NaD absorption line, the narrow strong line around 5900 \AA is stronger in the bulge than in the disk, 
consistent with the hypothesis that bulges could be Na-enhanced (e.g. in the bulge of the Milky Way, on average [Na/Fe] $\sim 0.2$, 
\citealt{Fulbright2007}). However, we do not use this particular line in the following analysis because 
its blue band is heavily contaminated from emission sky-lines and because of the recent finding presented in 
S+15 showing that the NaD lines are model dependent and therefore must be interpreted with caution. 
The aTiO index has also been excluded from the analysis because it is contaminated from emission sky-lines. 

\placefigstelpop

In Figure~\ref{fig:index-index} we show index-index plots of the selected optical indicators 
to provide qualitative constraints on the stellar population parameters. In panels a) and b) the IMF dependence is minimal 
and the age and abundances of the galaxy can be inferred. 
The IMF normalization of the galaxy can be instead inferred from panels c) and d), 
using SSP models with the metallicity and age previously determined. 

\subsection{Age and metallicity of the stellar population}
\label{subsect:agemetal}
In the upper panel of Figure~\ref{fig:index-index} we show index-index diagrams 
of some of the classical Lick-indices. 
These diagrams are widely used to constrain age and metallicity of
integrated stellar populations. 
The average indices extracted from the different regions are plotted 
on top of grids of CvD12 SSP models with varying ages (blue lines, [7,9,11,13.5] Gyr),
\alphafe\, (red lines, [0.0, 0.2, 0.4] dex) and total metallicity
(green lines, [-0.22, 0.0, +0.22]). All plotted models have a
Chabrier-like IMF, but for these indicators (\Hbeta, \Mgb,
Fe5270, Fe5335) the dependency on IMF is minimal.

Panel a) of Figure~\ref{fig:index-index} shows the \Hbeta--[MgFe] diagram.
The \Hbeta\, index is one of the most used age-indicators in the optical, and allows one to 
break the age-metallicity degeneracy,  when combined with the 
commonly-used [MgFe] index.   
However, this diagram alone does not allow one to also break degeneracies with \alphafe. 
This is clear for instance in the case of the disk component. From panel a) it is inferred that 
the bulge has an old population with solar \alphafe\, and slightly sub-solar metallicity. The central 
region is also old but has super-solar abundances. 
For the disk the interpretation is somewhat more difficult. In fact, the disk spectrum matches a model 
with age $=11 \pm 2$ Gyr, slightly super-solar metallicity and solar \alphafe\, but also a 
model with older age, lower metallicity and $\alpha$-enhanced. 
To discriminate between these two possibilities, the reader should refer to the results plotted in panel b).
Panel b) shows  ${\mathrm{Mg}b}$ versus $<Fe>$, which is the mean of Fe5270 and Fe5335. 
Here the disk matches only with a model with age $=11 \pm 1$ Gyr, solar metallicity and solar  \alphafe. 

Index-index plots are a useful tool to give qualitative inference 
on the stellar population parameters, but it is sometimes hard to break degeneracies 
and quantitatively constrain age, metal abundance and IMF slope at the same time.
Therefore, in order to give more quantitative constraints on the stellar population parameters of the three components, we perform a 
$\chi^{2}$ minimization comparing the 1D spectrum of each component with grids of linearly interpolated SSPs spanning a range of ages
(between 7 and 13.5 Gyr with steps of 1  Gyr,), \alphafe\, (between −0.2 and +0.4 dex, with a step of 0.1 dex), and total metallicity
([M/H]=[-0.22,0.0,+0.22])  for a Chabrier IMF (we note that a moderate change in the IMF slope has only a
negligible effect on  these blue features). 
We find that the disk is slightly younger and more metal-rich than the bulge, consistent with a SSP model of $\sim 10.5_{-2.0}^{+1.6}$ Gyr, solar
total metallicity ([M/H]=$-0.03^{+0.02}_{-0.1}$) and solar \alphafe (\alphafe$=+0.04^{+0.02}_{-0.01}$). 
The bulge fits better with a $13.5^{+1.4}_{-1.4}$ Gyr model with sub-solar metallicity
([M/H]$=-0.17^{+0.12}_{-0.1}$) and solar \alphafe (\alphafe$=+0.03^{+0.03}_{-0.01}$). 
The central region shows stellar population properties  in between those of the disk and
those of the bulge, as expected. It is as old as the bulge ($12.9^{+1.5}_{-1.8}$ Gyr), 
shows a solar or slightly super solar \alphafe\, (\alphafe$=+0.09^{+0.03}_{-0.01}$) 
and it is even more metal enriched ([M/H]=$+0.23^{+0.08}_{-0.05}$).
These SSP results nicely confirm the qualitative picture obtained via index-index plots in the upper 
row of Figure~\ref{fig:index-index}, and are summarized in Tab.~\ref{tab:SSP}. 
However, we note that since the 1D spectra of bulge and disk are composite spectra where
the contributions of the two components are not perfectly disentangled, the true stellar population differences are likely
to be even greater.

\placetab
Finally we consider NGC~4697 as whole and calculate gradients in its stellar populations.
Following the approach of \citet{Tortora2010b},  we define the stellar  population gradient 
as the dimensionless coefficient of the relation 
\begin{equation}
\log X -- \log R/R_{\rm eff}(\nabla_{X} = \frac{\delta (\log X)}{\delta \log (R/R_{\rm eff})}), 
\end{equation}
where X represents the age or the total  metallicity expressed in terms of [M/H].
We find a non negligible but very small age gradient and a  negative metallicity 
gradient in the sense that NGC~4697 is more metal-rich and older in the center. 
These gradients are consistent with the measured gradients in old early-type galaxies 
(\citealt{Kuntschner2010, Tortora2010b})  and with remnants of gas-rich major mergers 
from simulations (e.g. \citealt{Hopkins2009}, H+09). 

\subsection{The IMF normalization}
Using the metallicity and age constraints obtained from the Lick blue indices via the $\chi^{2}$ minimization, we focus now on two
of the IMF-sensitive optical indicators that fall in the VIMOS wavelength range, the bTiO and the TiO1. 
We plot them against the [Mg/Fe] index in the lower panels of Figure~\ref{fig:index-index}. The combination of the latter indicator with 
IMF-sensitive features breaks the degeneracies between the \alphafe, total metallicity and IMF slope variations. 

For each component,  we use SSP models with varying IMF slope (x between 1.8 and 2.5) and  \alphafe, and with the other stellar population
parameters as close as possible to the ones derived from the $\chi^{2}$ fit to the blue indices. For the bulge, we use SSP
models with 13.5 Gyr and [M/H] = -0.22, for the disk we use a SSP models with 11.0 Gyr and [M/H] = 0.0 (solar). 
Finally, for the mixed, central population we use SSP  models with 13.5 Gyr 
and [M/H] =+0.22 (super-solar).

In Figure~\ref{fig:index-index} black lines show models with different IMF slopes and red lines show models with varying  \alphafe. 
No detectable variation in the IMF normalization between bulge and disk is visible from the two panels. 
Both components of NGC~4697, as well as the  central region, are consistent with a normalization of $x=2.0 \pm 0.2$, 
slightly dwarf richer than a Chabrier IMF (x=1.8, still consistent within 1-sigma) but shallower than a Salpeter (x=2.35), 
which is perfectly consistent with the slope predicted from the IMF--$\sigma_{\star}$ relation (e.g.,
\citealt{LaBarbera2013, Tortora2013}; S+14) for an ETG with $\sigma_{\star} = 160$ \kms. 
IMF normalization does not change with radius for this galaxy is consistent with the results found in  \citet{MartinNavarro2014}. 
The lack of IMF spatial gradient is also consistent with the results on the low-mass system in 
\citet{MartinNavarro2014} that shows an IMF normalization constant with radius and consistent with a Kroupa-like profile.

\subsection{Linking kinematics and stellar populations }
The main conclusion from the stellar population analysis is that the inner region (typically $\sim $\Reff$/8$) 
looks coeval (or a bit older) and more metal-rich than the regions farther out, 
which suggests that this central region has been formed early via rapid
collapse. 
However, from the stellar population analysis alone, we cannot exclude minor or major merging of
late-type disk galaxies, which have destroyed spiral arm features
and left a younger disk component. It is therefore necessary to connect and 
correlate the stellar population results with those on the 
spatially resolved kinematics. 

The SAURON team (\citealt{Kuntschner2006, Krajnovic2008}) have found that 
the fast-rotating component almost always features a higher metallicity compared to the galaxy as a whole. 
They have attributed this to the presence of a later-forming stellar population in a rotating structure with disk-like 
kinematics (\citealt{Kuntschner2010}). 
From the NGC~4697 2D maps as well as from the stellar population analysis 
presented in this study, we infer the same scenario.
This is demonstrated inFigure~\ref{fig:h4hbeta}, which shows the 
correlation between the $h_{3}$ (top panels) and the $h_{4}$ (bottom panels) parameters 
and the \Hbeta\, (left) and \Mgb\, (middle) values obtained from the smoothed 2D maps. 
We run a Spearman test on these relations to better quantify the statistical dependence 
between the high-order moments of the LOSVD  and  \Hbeta\ and \Mgb\, respectively.  
The resulting  Spearman rank correlation coefficients are $\rho=0.36$ ($0.2$) and $\rho=0.32$
($0.36$) for $h_{4}$ ($h_{3}$) and \Hbeta\, and \Mgb\, respectively. Given the big sample size 
(more than 100 points), these values of $\rho$ all exceed the critical value ($\rho_{crit} (n>100) =0.165$) 
and therefore the relations are statistically significant, although not strong. 

The right panels of Figure~\ref{fig:h4hbeta} shows the \Hbeta\, --$h_{3}$ (top panel) 
and --$h_{4}$ (bottom panel) diagrams  
with the points color-coded according to the \Mgb\, values. 
Although the discussion must be kept on a qualitative level because 
of the well-known age-metallicity degeneracy (\citealt{Worthey1994}), 
we can simplify the situation and say that 
larger \Hbeta\, values indicate younger ages and larger \Mgb\, values indicate higher total metallicity. 
Thus we can conclude that points with $h_4>0$, tracing the later-formed disk,  
are slightly younger and more metal rich, while points with $h_4<0$, roughly tracing the bulge component 
because they are along the minor axis of NGC~4697,  all have generally older age and lower metallicity.
The same holds also for points with larger $|h_{3}|$, although in this case the relations are less robust.   

Remarkably, the spatially resolved stellar population results perfectly agree (on a qualitative level at least) 
with the spatially resolved kinematics results.
Unfortunately none of the hydrodynamical simulations derive 2D detailed stellar population properties. 
In the future, it might be extremely useful to have predictions from simulations on the ages and metal abundances
of the remnant of different mergers scenarios. 

In the SAURON ETGs of low to intermediate mass ($\sigma =100$--$160$ \kms), the region of increased metallicity is
connected to a mild decrease of the \alphafe, whereas in more massive galaxies ($\sigma > 160$ \kms) the disk formation and
increased metallicity is typically constrained to a central location involving only a minor fraction of the total mass of the
galaxy and only in about 25 per cent of the cases also a central depression in \alphafe\, can be seen. 

In the case of NGC~4697, we find that the central region is 
slightly $\alpha-$enhanced with respect to the bulge, whereas the disk shows roughly the same [$\alpha$/Fe] as the bulge.
A larger metallicity associated with the fast rotating component, seems to point toward 
a formation scenario where fast rotator galaxies have undergone one or multiple periods of secondary star-formation 
at a later stage and over longer timescales and therefore resulted in an increased metallicity and a decrease of the abundance ratio
in a disk-like structure (\citealt{Kuntschner2010}). 

A younger and more metal-rich disk is also well
consistent with the scenario suggested by the gas-rich merger (e.g. H+09), as the younger disk would arise when the gas, retaining
its angular momentum during the merger, is fueled in the center of the remnant and eventually turns into stars.  
In Section \ref{subsect:2Dmaps} we have demonstrated that this same scenario is also perfectly able to explain the kinematical
substructures. 

We note here that the disk merging scenario does not necessarily imply major merging, as also merging with lower mass
fraction can produce similar kinematical features. 
In fact, simulations in N+14 have shown that  correlations similar to the ones we observe in the $h_{3}$ and $h_{4}$ 
parameters of NGC~4697 can originate from two galaxy formation scenarios: a late gas-rich major merger 
or in situ star formation plus minor merging. 

\placefiglast

In conclusion, the combination of the kinematical features and the stellar population properties seem to point to the scenario where
some fast rotating and disky ellipticals can have formed from dissipative (wet) mergers of disk galaxies.

\section{Discussion and conclusions}
In this paper, we have presented high S/N integral field kinematics and stellar population 
combined analysis of the elliptical galaxy NGC~4697, 
an almost edge-on, intermediate luminosity ($L_ {*}$) isolated elliptical galaxy.
We used a mosaic approach,  the first of its kind with VIMOS IFU data, obtaining 
2D maps of stellar kinematics and absorption lines up to 0.7\Reff.
The final effective field-of-view of $\sim 80"\times 60"$ allowed up to obtain one among the largest 
2D kinematical maps ever measured for a single early-type galaxy (in terms of arcsec) 
 providing a unique view on the internal structure of the system, as well as crucial morphological 
and dynamical information. 

Our main scientific findings are the following:
\begin{itemize}
\item[i)] We have measured the 2D velocity moments of the galaxy's stellar population finding 
clear evidence for a rotation-supported structure along the major axis from the rotation 
velocity and the third Gauss-Hermite coefficient $h_{3}$ (representing the skewness). 
We have  measured the ``spin-parameter" (\lambdaR) confirming the classification 
of this system as `fast-rotator'. 
\item[ii)] We have analyzed the correlations between the  $h_{3}$ and  $h_{4}$ parameters 
with $V/\sigma$ (Fig.~\ref{fig:vsig}), together with the \lambdaR\, spatial profile, demonstrating how these characterize the 
different kinematical components (bulge and disk) of the galaxy.  
\item[iii)] We have directly compared these correlations with those predicted by hydrodynamical 
simulations of  \citet{Hoffman2009} from gas-rich disk-disk major mergers with a 30\% gas fraction and of 
\citet{Naab2014} for gas-rich minor and major mergers (class A and B).
The correlations we presented  closely resemble the theoretical predictions and indicate that the mass assembly of NGC~4697
is dominated by gas-rich minor mergers, gradual dissipation plus possibly a late gas-rich major merger. 
\item[iv)] A direct comparison with 1D long-slit spectroscopy (\citealt{deLorenzi2008})  
shows an overall good agreement between the different kinematics results along the major axis. 
\item[v)] We extracted 1D spectra of the disk and the bulge of NGC~4697, using 
a two-components fit to a deep V-band image. 
\item[vi)] We studied the stellar populations of these two components using SSP models. 
We find indications  that the stars in the disk are $\sim 2-3$ Gyr younger and 
more metal rich that the stars in the bulge.  
\item[vii)] We linked the stellar population analysis and the internal kinematics of the system by correlating 
age and metallicity of the stellar populations of the two components with the $h_{3}$ and  
the $h_{4}$ kinematic parameters, finding further evidence for a  significant in-situ formation of stars since $z\sim 2$ 
with additional gas-rich minor mergers (although we do not exclude the possibility of a late major-merger that 
left a previously rapidly rotating system unchanged). 
\item[viii)] From the two bluer IMF-sensitive optical indicators of  S+14, which fall in the VIMOS wavelength range,
we do not detect any difference in the IMF slope between the bulge and the disk. 
Both components are consistent with an IMF with a single power slope of $x$=2.0, consistent with a 
Kroupa IMF and slightly lighter than a Salpeter-like IMF. The inferred IMF slope is perfectly consistent with 
the one predicted from the IMF--$\sigma_{\star}$ relation for an  ETG with $\sigma_{\star} = 160$ \kms, such as 
NGC~4697 (\citealt{Conroy2012b, LaBarbera2013, Treu2010, Tortora2013, Tortora2014a}, S+14). 
\end{itemize}

The absolute novelty of this work resides: i) on the depth and width of the data acquired as we have very high $S/N$ 
spectra over a  an effective area of the sky of $80''\times60''$ (e.g. larger than MUSE FOV); ii) on the presentation of a detailed spatially 
resolved kinematical and population 2D analysis within the central effective radius, which has allowed us to cross-correlate LOSVD
 information (i.e. $h_3$ and $h_4$) with relevant line indices (e.g. $h_4$ vs. \Hbeta\, and \Mgb\, in  Figure~\ref{fig:h4hbeta}), showing that points 
 with $h_4>0$, associate mainly to the later-formed disk,  are younger and more metal rich, while 
 points with $h_4<0$, roughly tracing the bulge component because aligned with the minor axis of NGC~4697,  
 all have generally a  smaller \Hbeta\, and smaller [M/H].

The emerging picture is that a later formed disk has been built from pre-enriched gas that formed stars after having 
dissipatively sunk into the center. This is well consistent with the scenario suggested by 
gas-rich disk merging simulations (e.g. H+09), as the younger disk would arise when the gas, 
retaining its angular momentum during the merger, is fueled in the center of the remnant and eventually turns into stars.  
It is remarkable that within this scenario, photometry, stellar population features and kinematical features like 
the $h_3$- and $h_4$-$V/\sigma$ correlations are 
all qualitatively correctly predicted.  We stress again that H+09 only considers 1:1 disk-disk merger but that also 
merging with lower mass fraction (minor mergers) can produce similar kinematical features (e.g. N+14). 
Furthermore other mechanisms involving a rapid dissipative collapse of enriched gas 
(clumpy disks, violent relaxation, cold accretion), might have produced a similar galaxy component.  

This paper is intended as a pilot program for a large project that  
aims at the study of internal structure, mass distribution and stellar population of a sample of nearby $L_ {*}$ ETGs.
With medium-high signal-to-noise spatially resolved spectra, we will perform a detailed 2D study 
of the stellar population via line-index measurements and try to further test possible variation in the IMF normalization. 
We will attempt to characterize the low-mass end of  Initial Mass Function normalization in the bulge and in the disk components 
by measuring EWs of the IMF-sensitive features defined in S+14. In order to achieve this result we plan to change 
the configuration of VIMOS in the new observations, choosing a redder filter (HR-red) to cover the proper wavelength range\footnote{We are aware that 
moving to the red filter will cause the loss of some of the blue indices covered in this paper, making age determination more complex. We therefore plan to select 
ETGs where measurements of age and metallicity are available from the literature and mainly focus on measuring gravity-sensitive features 
to study the possible spatially variation of the IMF slope, with particular regard on bulge versus disk dominated zones.}. 
In addition to a direct measurement  of the stellar mass-to-light ratios (M$^{\star}$/L), thanks to the 2D IF coverage, 
we will be able to study age and metallicity  radial gradients for our systems. 
Moreover, where spatial gradients in the stellar population parameters are present, the assumption of a spatially-constant (M$^{\star}$/L)
clearly does not work. Consequently, the 2D coverage, allowing a more accurate measure of the (M$^{\star}$/L), also permits a more accurate and 
complete dynamical analysis (e.g., \citealt{Tortora2010,Tortora2011}).  

A quantitative dynamical model to fit the whole 2D maps for the rotation 
and the velocity dispersion, is out of the purpose of this paper but will be performed in future work.  
Only with a complete 2D kinematical map of a galaxy (up to at least $1$\Reff) as well as a robust dynamical model it 
will be possible to measure the DM fraction, trace its radial distribution as a function of the mass of the halo, 
and test plausible environment dependencies.
In future work we therefore plan to perform a detailed study of the stellar content of ETGs to 
combine spatially resolved stellar information on the central region with planetary nebula  
data on the outskirts (up to $\sim 5'$) with the final purpose of obtaining the complete rotation curve and 
distribution of dark matter (e.g. \citealt{Napolitano2009,Napolitano2011}).  

Moreover, thanks to a complete and detailed dynamical analysis, we will have a completely independent measurement of the IMF slope that 
we will compare to the one obtained by stellar population modeling. If both methods agree on the resulting IMF, this will be 
a strong test of the absence of hidden systematics.  

Finally another possible interesting follow-up will be studying the dynamics and the IMF normalization of this galaxy in the phenomenological 
framework of  the Modified Newtonian Dynamics theory (MOND, \citealt{Milgrom1983a,Milgrom1983b}, following the approach presented in \citealt{Tortora2014a}).

\section*{Acknowledgments}
The use of the Penalized Pixel Fitting developed by Cappellari \& Emsellem is gratefully acknowledged. 
Data were reduced using EsoRex and XSH pipeline by ESO Data Flow System Group.
C.S. thanks S.C. Trager for the use of the code SPINDEX2 and C. Conroy for the precious help in 
understanding and using his stellar population models. 
C.T. has received funding from the European Union Seventh Framework Programme (FP7/2007-2013) under grant agreement n. 267251. 
We thank the referee for very constructive comments.
%


\end{document}